\documentclass[%
 aip,
 amsmath,amssymb,
 reprint,%
]{revtex4-1}

\usepackage{graphicx}% Include figure files
\usepackage{dcolumn}% Align table columns on decimal point
\usepackage{bm}% bold math
\usepackage[utf8]{inputenc}
\usepackage[T1]{fontenc}
\usepackage{mathptmx}
\usepackage{etoolbox}
\usepackage{xcolor}
\usepackage{amssymb}

\makeatletter
\def\@email#1#2{%
 \endgroup
 \patchcmd{\titleblock@produce}
  {\frontmatter@RRAPformat}
  {\frontmatter@RRAPformat{\produce@RRAP{*#1\href{mailto:#2}{#2}}}\frontmatter@RRAPformat}
  {}{}
}%
\makeatother
\begin{document}

\preprint{AIP/123-QED}

\title[]{Exact moment equivalence and structural nonidentifiability in nonlinear epidemics}
% Force line breaks with \\
\author{Roni Muslim}
\email{roni.muslim@apctp.org}
\affiliation{ 
Asia Pacific Center for Theoretical Physics,  Pohang, 37673, Republic of Korea
}%
\affiliation{ 
Research Center for Quantum Physics, BRIN, South Tangerang, 15314, Indonesia
}%
%  \altaffiliation[Also at]{Asia Pacific Center for Theoretical Physics,  Pohang, 37673, Republic of Korea}%Lines break automatically or can be forced with \\
% \author{B. Author}%
%  \email{Second.Author@institution.edu.}
% \affiliation{ 
% Authors' institution and/or address%\\This line break forced with \textbackslash\textbackslash
% }%

% \author{C. Author}
%  \homepage{http://www.Second.institution.edu/~Charlie.Author.}
% \affiliation{%
% Second institution and/or address%\\This line break forced% with \\
% }%

\date{\today}% It is always \today, today,

\begin{abstract}
Transmission through temporary groups does not necessarily preserve complete
information about the group-size distribution in aggregate epidemic data. We
show that, in finite-population SIS and SIR models, the group-size distribution
enters the dynamics only through a finite set of moments selected by the order
of the nonlinear transmission kernel. Consequently, markedly different
distributions can generate identical stochastic dynamics when their relevant
moments coincide. This equivalence extends to transient evolution,
fluctuations, extinction-time statistics, and final outbreak sizes. In the
deterministic limit, the first moment sets the invasion threshold, whereas the
second controls the nature of the transition and the emergence of bistability
and hysteresis. The two distributions become distinguishable only when a
higher-order transmission mechanism activates their first unmatched moment;
even a weak additional channel can shift the phase boundary and place the
systems in different dynamical regimes. Solutions of the master equation and
stochastic simulations support these analytical predictions. These results
establish an intrinsic limit on epidemic inference: a single aggregate
dynamical protocol can identify only an equivalence class of group-size
distributions, rather than uniquely reconstructing the full distribution.
\end{abstract}

\maketitle

\begin{quotation}
Disease transmission often occurs in temporary groups whose sizes vary from
one encounter to the next, yet aggregate epidemic data retain only partial
information about these interaction patterns. For nonlinear SIS and SIR
processes in randomly assembled groups, we find that the entire influence of
the group-size distribution reduces to a finite set of factorial moments
selected by the order of the transmission mechanism. Group-size distributions
with markedly different forms can therefore generate exactly the same
stochastic epidemic dynamics, even in finite populations. This moment
hierarchy also organizes the nonlinear phase structure: the first moment sets
the invasion threshold, whereas the second controls the emergence of
bistability and hysteresis. Differences hidden at a given transmission order
become observable only when a higher-order channel activates the next
unmatched moment, allowing two previously equivalent systems to enter
different dynamical regimes. These results connect higher-order epidemic
interactions with bifurcation structure and clarify which features of group
organization can, and cannot, be recovered from aggregate observations.
\end{quotation}

\section{Introduction}
\label{sec:introduction}

Compartmental models such as SIS and SIR provide a compact mathematical
framework for relating transmission and recovery to changes in the numbers
of individuals in different epidemiological states
\cite{kermack1927,hethcote2000}. When contact patterns are included,
network-based approaches show that epidemic thresholds, outbreak sizes, and
disease persistence can be altered by degree heterogeneity, contact
clustering, and correlations between individuals
\cite{newman2002,keeling2005,pastor2015}. Real contacts, however, are not
always persistent. Encounters form, change, and dissolve, so the timing and
duration of interactions can affect transmission pathways even when the
corresponding aggregate networks are identical
\cite{holme2012,perra2012,valdano2015,masuda2017}. Despite these advances,
most network models still resolve interactions into pairs. Such a reduction
can discard relevant information when exposure occurs simultaneously within
households, classrooms, meetings, public transportation, or social
gatherings.

These settings are more naturally described by higher-order interactions, in
which a single event involves three or more individuals. Hypergraphs and
simplicial complexes preserve the identity of a group instead of replacing it
with a collection of pairwise edges
\cite{lambiotte2019,battiston2020,battiston2021,boccaletti2023}. This
distinction is not merely representational. Two structures with the same
network projection can support different collective dynamics because their
group compositions and group sizes differ. Activity-driven models,
statistical filtering of hyperedges, and motif-based analyses have shown that
group size, recurrence, and internal organization contain information that
cannot be recovered from a pairwise network
\cite{petri2018,musciotto2021,lotito2022}. Temporal interaction data further
reveal memory and correlations between groups of different sizes
\cite{gallo2024}. These findings make the group-size distribution part of the
transmission mechanism rather than an ancillary property of contact data.

The dynamical consequences of group interactions become especially
pronounced when the transmission rate is nonlinear in the number of infected
individuals within a group. Simultaneous exposure to several infectious
sources can produce synergistic effects, superspreading events, or a rise in
risk that is faster than the sum of independent contact contributions.
Contagion models on simplicial complexes and hypergraphs show that such
mechanisms can turn a continuous epidemic transition into a discontinuous
one, create bistability and hysteresis, and introduce a critical seed size
\cite{iacopini2019,dearruda2020,matamalas2020,landry2020}. The infection
kernel, membership heterogeneity, group-size distribution, and temporal
organization of interactions can each shift epidemic thresholds or change
the bifurcation structure
\cite{stonge2021,chowdhary2021,stonge2022,higham2022,lucas2023}. More recent
work has shown that these outcomes also depend on the chosen representation,
overlap between interactions, the presence of higher-order components, and
correlations in participation across groups of different sizes
\cite{zhang2023,burgio2024,kim2024,malizia2025overlap,
malizia2025heterogeneity,guzman2026}. The group-size distribution and its
microscopic organization can therefore influence epidemic phase behavior
through several distinct routes.

Although group heterogeneity has received considerable attention, a more
basic question remains unresolved: which features of the group-size
distribution can actually be read from aggregate epidemic dynamics? Many
previous analyses fix a small number of interaction orders, retain quenched
hyperedges, or rely on mean-field closures and approximate master equations.
These approaches are useful for calculating thresholds and stationary
states, but they do not by themselves establish whether different group-size
distributions can generate the same stochastic process. This question must
be distinguished from agreement in the deterministic limit. In finite
populations, fluctuations, absorbing states, and extinction-time
distributions are integral parts of the dynamics
\cite{gillespie1977,nasell1999,dickman2002}. Agreement in an epidemic
threshold or stationary prevalence is therefore insufficient to establish
equivalence between two group-interaction mechanisms. A comparison is needed
at the level of the Markov generator, covering every transition rate at every
admissible population state.

This issue is directly related to identifiability. A model is structurally
nonidentifiable when distinct mechanisms or parameter values produce the
same ideal output, even with continuous and noise-free observations
\cite{bellman1970,miao2011,villaverde2019}. In epidemic models,
identifiability also depends on which quantities are observed, such as
prevalence, incidence, or combinations of several outputs
\cite{evans2005,dankwa2022}. This differs from practical
nonidentifiability, which arises from measurement noise, limited
observations, or weak output sensitivity
\cite{wieland2021,heinrich2025}. Additional experiments or observables can
reduce degeneracy at the output level
\cite{ovchinnikov2022,bortner2024}, but they cannot distinguish two models
with identical stochastic generators. For group interactions, the object to
be inferred is not merely a collection of scalar parameters but the
distribution $Q(n)$ itself. It is therefore necessary to determine whether
the dynamics contain information about the full distribution or only about
a finite set of its moment combinations.

In this work, we address this question for SIS and SIR models with temporary
groups that are independently reassembled at every exposure event. Group
composition is averaged exactly using the hypergeometric distribution,
without a binomial approximation or mean-field closure. For a polynomial
transmission kernel of order $q$, we prove that the finite-population Markov
generator depends on the group-size distribution only through $q$
generalized factorial moments. Consequently, distinct distributions,
including distributions with disjoint supports, generate the same law for
the aggregate process whenever these moments coincide. We then derive the
deterministic limit and show that the first moment sets the invasion
threshold, whereas the second controls the direction of the bifurcation, the
emergence of bistability, and the width of the hysteresis region. This moment
hierarchy is examined through solutions of the master equation and
stochastic simulations for both SIS and SIR dynamics. Finally, we activate
one additional higher-order transmission channel to show how the first
unmatched moment breaks the equivalence and shifts the saddle-node
threshold. The resulting framework places a precise limit on inference from
aggregate epidemic data: a single nonlinear transmission protocol generally
identifies an equivalence class of group-size distributions rather than the
full distribution itself.

\section{Model description}
\label{sec:model}

We consider a well-mixed population of fixed size $N$, in which each
individual is susceptible ($S$), infected ($I$), or recovered ($R$). The
variables $S(t)$, $I(t)$, and $R(t)$ denote the numbers of individuals in
the corresponding epidemiological states at time $t$. Transmission occurs
through temporary groups whose sizes and memberships are independently
resampled at every infection opportunity. The interaction structure is
therefore annealed: no permanent hyperedges, persistent
groups, or correlations between group members are retained from one event
to the next. This formulation serves as a reference model without
topological correlations, allowing the effect of the group-size distribution
on epidemic dynamics to be isolated directly.

The dynamics is defined as a continuous-time Markov process with
focal-centered infection events. The instantaneous infection hazard of each
susceptible individual is obtained by averaging the conditional infection
rate over all temporary groups that the individual may experience. To
construct this hazard microscopically, consider a susceptible focal
individual and draw a group size $n$, with
$2\leq n\leq n_{\max}\leq N$, from the participant-view distribution
$Q(n)$. The remaining $n-1$ group members are then sampled uniformly,
without replacement, from the other $N-1$ individuals. Only the focal
individual can undergo the transition $S\to I$ during a single infection
event, so the number of infected individuals can increase by at most one
per event. The parameter $\beta$, introduced below, incorporates the
baseline frequency of exposure opportunities and the efficiency of
transmission.

If $P(n)$ denotes the group-size distribution at the event level, the size
distribution experienced by a randomly selected participant is the
size-biased distribution
$Q(n)=nP(n)/\langle n\rangle_P$, where
$\langle n\rangle_P=\sum_n nP(n)$. Both distributions are normalized,
$\sum_{n=2}^{n_{\max}}P(n)
=\sum_{n=2}^{n_{\max}}Q(n)=1$. The size bias arises because larger groups
contain more participants and are therefore more likely to be experienced
by a randomly selected individual. We assume that participation is
independent of epidemiological state and remains unchanged throughout the
evolution. The same distribution $Q(n)$ consequently applies when the focal
individual is drawn from the susceptible subpopulation. While $P(n)$
describes group events, the dynamics below is defined from the perspective
of a focal participant and does not update all susceptible members of a
group simultaneously.

\begin{figure*}[tb]
\centering
\includegraphics[width=\linewidth]
{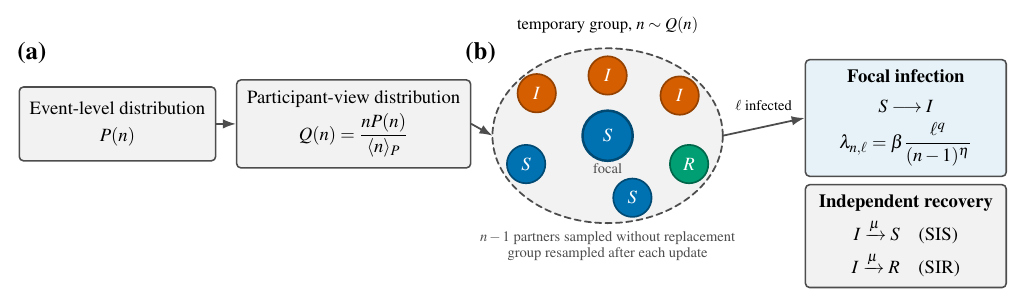}
\caption{Schematic illustration of the epidemic model with temporary groups.
(a) The event-level group-size distribution, $P(n)$, is converted into the
participant-view distribution,
$Q(n)=nP(n)/\langle n\rangle_P$. (b) A focal individual interacts with
$n-1$ partners sampled without replacement. If $\ell$ of these partners are
infected, the focal individual becomes infected at rate
$\lambda_{n,\ell}$, whereas recovery occurs independently at rate $\mu$.
The group is resampled after every update.}
\label{fig:model_schematic}
\end{figure*}

Suppose that the population contains $I$ infected individuals. Because the
focal individual is susceptible, all $I$ infected individuals belong to the
remaining population of size $N-1$. Conditional on a group size $n$, the
number $\ell$ of infected individuals among the $n-1$ partners follows the
hypergeometric distribution
\begin{equation}
\Pr(\ell\mid n,I)
=
\frac{
\binom{I}{\ell}
\binom{N-1-I}{n-1-\ell}
}{
\binom{N-1}{n-1}
},
\label{eq:group_composition}
\end{equation}
with support
$\ell_{\min}=\max(0,n-N+I)$ and
$\ell_{\max}=\min(I,n-1)$. The lower bound accounts for cases in which a
sample of size $n-1$ must contain a minimum number of infected individuals
because too few uninfected individuals are available; the upper bound is
set by the number of infected individuals in the population and the sample
size. Equation~\eqref{eq:group_composition} retains finite-population
fluctuations in group composition exactly, without a binomial approximation
or mean-field assumption. The group-formation protocol, focal update, and
recovery channels are summarized schematically in
Fig.~\ref{fig:model_schematic}.

If the temporary group experienced by a focal susceptible contains $\ell$
infected individuals among its $n-1$ partners, the conditional infection
rate of the focal individual is defined as
\begin{equation}
\lambda_{n,\ell}
=
\beta\frac{\ell^q}{(n-1)^\eta},
\label{eq:infection_rate}
\end{equation}
where $\beta>0$ is the baseline transmission scale and has dimensions of
inverse time, $q\in\mathbb{N}^{+}$ is the order of the nonlinearity, and
$\eta\geq0$ controls the dependence of exposure on group size. For $q=1$,
the contributions of infected partners are additive. By contrast, $q>1$
represents synergistic amplification, in which the simultaneous presence of
several infected individuals produces an infection rate that grows
nonlinearly with $\ell$. The choice $\eta=0$ corresponds to exposure that
is not normalized by group size, whereas $\eta=1$ removes one factor of
$n-1$. More generally, $\eta=q$ gives a prevalence-based kernel,
$\lambda_{n,\ell}=\beta[\ell/(n-1)]^q$. When $\ell=0$,
$\lambda_{n,0}=0$, so a group containing no infected partners cannot produce
transmission. The polynomial dependence on $\ell$ provides a simple
representation of synergistic transmission while also allowing the
microscopic kernel to be related exactly to a finite set of factorial
moments of the group-size distribution.

For a population state containing $I$ infected individuals, the effective
infection hazard experienced by one susceptible individual is obtained by
averaging Eq.~\eqref{eq:infection_rate} over all possible group sizes and
compositions:
\begin{equation}
\overline{\lambda}(I)
=
\sum_{n=2}^{n_{\max}}Q(n)
\sum_{\ell=\ell_{\min}}^{\ell_{\max}}
\Pr(\ell\mid n,I)\lambda_{n,\ell}.
\label{eq:individual_averaged_hazard}
\end{equation}
Each pair $(n,\ell)$ can be viewed as an exposure channel contributing
$Q(n)\Pr(\ell\mid n,I)\lambda_{n,\ell}$ to the infection hazard of the
focal individual. Equation~\eqref{eq:individual_averaged_hazard} therefore
accounts exactly for fluctuations in both group size and composition
without introducing simultaneous updates of other group members. Because
all susceptible individuals are exchangeable and experience the same
sampling mechanism, the aggregate infection rate is
$W^{\mathrm{inf}}(S,I)=S\overline{\lambda}(I)$. For the SIS model,
$S=N-I$, and the transition $I\to I+1$ occurs at rate
$W_I^{+}=(N-I)\overline{\lambda}(I)$. For the SIR model, the transition
$(S,I)\to(S-1,I+1)$ occurs at rate
$W_{S,I}^{\mathrm{inf}}=S\overline{\lambda}(I)$. Thus, a single infection
event changes only the state of the focal susceptible, although its rate is
determined by the size and composition of the temporary group it
experiences.

Each infected individual recovers independently at rate $\mu$. In the SIS
model, recovery returns an infected individual to the susceptible state,
whereas in the SIR model it produces the transition $I\to R$, after which
reinfection is not allowed. The aggregate recovery rate is $\mu I$ in both
models. The macroscopic transitions and their rates can therefore be
summarized as
\begin{equation}
\begin{aligned}
\text{SIS:}\qquad
I&\longrightarrow I+1
&&\text{at rate }
(N-I)\overline{\lambda}(I),\\
I&\longrightarrow I-1
&&\text{at rate }
\mu I,\\[2mm]
\text{SIR:}\qquad
(S,I)&\longrightarrow(S-1,I+1)
&&\text{at rate }
S\overline{\lambda}(I),\\
(S,I)&\longrightarrow(S,I-1)
&&\text{at rate }
\mu I.
\end{aligned}
\label{eq:model_transition_summary}
\end{equation}
Because groups are independently reassembled at every interaction
opportunity, only the focal individual changes state during an infection
event, and individuals in the same epidemiological state are exchangeable,
$I$ in the SIS model and $(S,I)$ in the SIR model form closed Markov
processes. We define the population fractions as $s=S/N$, $i=I/N$, and
$r=R/N$, so that $s+i=1$ for SIS and $s+i+r=1$ for SIR. Group
configurations that do not cause infection generate no off-diagonal
transition in the Markov generator and therefore leave the macroscopic
state unchanged.

To characterize the information about the group-size distribution that can
enter the epidemic dynamics, we define the weighted factorial moments
\begin{equation}
M_k^{(\eta)}
=
\sum_{n=k+1}^{n_{\max}}
Q(n)
\frac{(n-1)_{\underline{k}}}{(n-1)^\eta},
\label{eq:generalized_moments}
\end{equation}
where $k=1,\ldots,q$ and
$(x)_{\underline{k}}=x(x-1)\cdots(x-k+1)$ denotes the falling factorial.
As shown in Subsec.~\ref{sec:exact_reduction}, after group composition is
averaged using Eq.~\eqref{eq:group_composition}, the aggregate Markov
generator depends on $Q(n)$ only through the vector
$\boldsymbol{M}_q^{(\eta)}
=(M_1^{(\eta)},\ldots,M_q^{(\eta)})$. This reduction is exact for finite
$N$ and is not a mean-field closure. Hence, for the same values of
$N$, $\beta$, $\mu$, $q$, and $\eta$, two group-size distributions with
identical $\boldsymbol{M}_q^{(\eta)}$ generate the same Markov generator.
If their initial conditions also coincide, the two processes have identical
aggregate path laws, including the mean prevalence, fluctuations, survival
probability, extinction-time distribution, and final outbreak-size
distribution. This result defines dynamically equivalent classes of
group-size distributions and establishes a structural limit on the
information about $Q(n)$ that can be recovered from aggregate epidemic
observations.

The analytical results developed below are tested using two complementary
numerical procedures. The finite-state master equation is solved to obtain
the state distribution, mean prevalence, survival probability, and final
outbreak-size distribution, while the Gillespie algorithm provides a
stochastic check that follows the microscopic update rules directly.
Details of the generator construction, master-equation integration,
calculation of SIR absorption probabilities, Gillespie implementation~\cite{gillespie1977},
parameter choices, number of realizations, and statistical uncertainty are
given in Appendix~\ref{app:numerical_procedures}.

\section{Results and discussion}
\label{sec:results}

\subsection{Exact reduction and the moment-sufficiency theorem}
\label{sec:exact_reduction}

We begin by averaging the infection rate over all possible group
compositions. When the population contains $I$ infected individuals, the
number $\ell$ of infected individuals among the $n-1$ partners of the focal
individual follows the hypergeometric distribution in
Eq.~\eqref{eq:group_composition}. Its $k$th factorial moment is
\begin{equation}
\mathbb{E}
\left[
(\ell)_{\underline{k}}
\mid n,I
\right]
=
\frac{
(n-1)_{\underline{k}}(I)_{\underline{k}}
}{
(N-1)_{\underline{k}}
},
\label{eq:hypergeometric_factorial_moment}
\end{equation}
where
$(\ell)_{\underline{k}}=\ell(\ell-1)\cdots(\ell-k+1)$. This identity
retains the finite-population correction arising from sampling group members
without replacement. Combinatorially,
$(n-1)_{\underline{k}}$ counts the ordered selections of $k$ distinct
positions within the group, whereas
$(I)_{\underline{k}}/(N-1)_{\underline{k}}$ is the probability that all
these positions are occupied by infected individuals.

Because $q$ is a positive integer, the ordinary power $\ell^q$ can be
expanded exactly in the falling-factorial basis as
$\ell^q=\sum_{k=1}^{q}\mathsf{S}(q,k)(\ell)_{\underline{k}}$, where
$\mathsf{S}(q,k)$ denotes a Stirling number of the second kind. Substituting
this identity and Eq.~\eqref{eq:hypergeometric_factorial_moment} into the
average infection rate gives
\begin{align}
\overline{\lambda}(I)
&=
\sum_{n=2}^{n_{\max}}Q(n)
\sum_{\ell=\ell_{\min}}^{\ell_{\max}}
\Pr(\ell\mid n,I)\lambda_{n,\ell}
\nonumber\\
&=
\beta
\sum_{k=1}^{q}
\mathsf{S}(q,k)
\frac{(I)_{\underline{k}}}
     {(N-1)_{\underline{k}}}
\sum_{n=k+1}^{n_{\max}}
Q(n)
\frac{(n-1)_{\underline{k}}}
     {(n-1)^\eta}
\nonumber\\
&=
\beta
\sum_{k=1}^{q}
\mathsf{S}(q,k)
M_k^{(\eta)}
\frac{(I)_{\underline{k}}}
     {(N-1)_{\underline{k}}}.
\label{eq:averaged_infection_rate}
\end{align}
In the second line, the average over group composition has been replaced by
the corresponding hypergeometric factorial moment, after which the order of
summation over $n$ and $k$ has been exchanged. The lower limit $n=k+1$
appears because $(n-1)_{\underline{k}}=0$ whenever $n\leq k$. Equation~\eqref{eq:averaged_infection_rate} is exact for a finite
population. Its derivation does not use the approximation
$\ell\approx(n-1)I/N$, a binomial distribution for group composition, or a
mean-field closure. The full distribution $Q(n)$ enters the infection rate
only through $M_1^{(\eta)},\ldots,M_q^{(\eta)}$. In the finite-population
analysis below, we assume $1\leq q\leq N-1$ so that all factorial
denominators are well defined. If $k>n_{\max}-1$, no group is large enough
to contain $k$ distinct positions, and therefore
$M_k^{(\eta)}=0$.

For the SIS model, the macroscopic state is completely specified by $I$
because $S=N-I$. The transition rates are
$W_I^+=(N-I)\overline{\lambda}(I)$ for infection and
$W_I^-=\mu I$ for recovery. The SIS generator acting on a function $f(I)$
is therefore
\begin{equation}
\left(\mathcal{L}_{\mathrm{SIS}}f\right)(I)
=
W_I^{+}\bigl[f(I+1)-f(I)\bigr]
+
W_I^{-}\bigl[f(I-1)-f(I)\bigr].
\label{eq:sis_generator}
\end{equation}
The state $I=0$ is absorbing because both transition rates vanish there. For the SIR model, the macroscopic state is represented by the pair $(S,I)$,
with $R=N-S-I$. The infection and recovery rates are
$W_{S,I}^{\mathrm{inf}}=S\overline{\lambda}(I)$ and
$W_{S,I}^{\mathrm{rec}}=\mu I$, respectively. The corresponding generator
is
\begin{align}
\left(\mathcal{L}_{\mathrm{SIR}}f\right)(S,I)
={}&
W_{S,I}^{\mathrm{inf}}
\bigl[f(S-1,I+1)-f(S,I)\bigr]
\nonumber\\
&+
W_{S,I}^{\mathrm{rec}}
\bigl[f(S,I-1)-f(S,I)\bigr].
\label{eq:sir_generator}
\end{align}
These two generators show that the numbers of individuals in the
epidemiological states form closed Markov processes. Information about
groups formed during previous events need not be retained because every
group is independently resampled.

\medskip
\noindent\textbf{Theorem 1 (factorial-moment sufficiency).}
Consider two group-size distributions $Q_a(n)$ and $Q_b(n)$ with the same
values of $N$, $\beta$, $\mu$, $q$, and $\eta$. If
$M_{k,a}^{(\eta)}=M_{k,b}^{(\eta)}$ for every $k=1,\ldots,q$, then the two
distributions generate identical aggregate Markov generators for both the
SIS and SIR models. Given the same initial condition, the probability
distributions of the corresponding population-count processes are identical
at every time $t\geq0$.

\smallskip
\noindent\textit{Proof.}
According to Eq.~\eqref{eq:averaged_infection_rate}, $Q(n)$ affects the
infection rate only through
$M_1^{(\eta)},\ldots,M_q^{(\eta)}$. Equality of these moments implies
$\overline{\lambda}_a(I)=\overline{\lambda}_b(I)$ for every admissible
value of $I$. The recovery rate $\mu I$ is independent of $Q(n)$, so all
SIS and SIR transition rates also coincide. The two processes therefore
have the same generators and Kolmogorov equations. Because their state
spaces are finite, the solutions of the Kolmogorov equations are unique.
Hence, when initialized from the same state, the probability distributions
of the two processes remain identical at all times.
\hfill$\square$

This theorem motivates the dynamic-equivalence relation
$Q_a\sim_q Q_b$, defined by
$M_{k,a}^{(\eta)}=M_{k,b}^{(\eta)}$ for every $k=1,\ldots,q$. A single
equivalence class may contain distributions with different supports,
shapes, variances, and tail behavior. Such differences cannot be detected
by observables that depend only on the aggregate process, including mean
prevalence, fluctuations in the infected count, survival probability,
extinction-time distributions, and final outbreak sizes. This statement
does not imply that the sequences of sampled groups or the individual
contact histories are identical; it is the probability law of the
population-count process that is the same.

The equivalence hierarchy is determined by the first unmatched moment.
Suppose that two distributions have identical moments up to order $m-1$ but
$\Delta M_m^{(\eta)}\neq0$, where
$\Delta M_k^{(\eta)}
=M_{k,a}^{(\eta)}-M_{k,b}^{(\eta)}$ and $m\leq q$. Defining
$\Delta\overline{\lambda}(I)
=\overline{\lambda}_a(I)-\overline{\lambda}_b(I)$ gives
\begin{equation}
\Delta\overline{\lambda}(I)
=
\beta
\sum_{k=m}^{q}
\mathsf{S}(q,k)
\Delta M_k^{(\eta)}
\frac{(I)_{\underline{k}}}
     {(N-1)_{\underline{k}}}.
\label{eq:rate_difference}
\end{equation}
For $I<m$, every term in this expression vanishes. The first difference
appears at $I=m$, for which
$\Delta\overline{\lambda}(m)
=\beta\mathsf{S}(q,m)\Delta M_m^{(\eta)}
m!/(N-1)_{\underline{m}}$. Thus, the first unmatched moment determines the
lowest factorial order at which the two processes can be distinguished.

This result establishes a limit on inference from aggregate epidemic data.
If $\beta$, $\mu$, $q$, and $\eta$ are known, time series of $S(t)$,
$I(t)$, and $R(t)$ can provide information at most about the vector
$\boldsymbol{M}_q^{(\eta)}
=(M_1^{(\eta)},\ldots,M_q^{(\eta)})$, rather than the full form of $Q(n)$.
If $\beta$ must also be estimated, the data identify only combinations of
$\beta$ and these moments. All distributions within the same equivalence
class assign identical probabilities to aggregate observations. The
inability to distinguish moment-matched distributions is therefore not a
consequence of insufficient data or limited numerical accuracy, but a
structural property of the model.

\subsection{Deterministic limit and epidemic phase structure}
\label{sec:deterministic_limit}

The macroscopic dynamics is obtained by taking the limit
$N\rightarrow\infty$ while keeping the infected fraction $i=I/N$ fixed.
This limit is taken for fixed $q$ and a group-size distribution with finite
factorial moments; group sizes are not allowed to grow proportionally with
$N$. For every fixed $k$,
$I_{\underline{k}}/(N-1)_{\underline{k}}\rightarrow i^k$. The contribution
of group interactions to the infection rate can then be summarized by the
effective exposure function
\begin{equation}
\Phi_q(i)
=
\sum_{k=1}^{q}
\mathsf{S}(q,k)M_k^{(\eta)}i^k,
\label{eq:effective_exposure}
\end{equation}
where $\mathsf{S}(q,k)$ denotes a Stirling number of the second kind. The
deterministic SIS equation becomes
\begin{equation}
\frac{di}{dt}
=
\beta(1-i)\Phi_q(i)-\mu i,
\label{eq:deterministic_sis}
\end{equation}
whereas the SIR model gives
\begin{equation}
\frac{ds}{dt}
=
-\beta s\Phi_q(i),
\quad
\frac{di}{dt}
=
\beta s\Phi_q(i)-\mu i,
\quad
\frac{dr}{dt}
=
\mu i.
\label{eq:deterministic_sir}
\end{equation}
The moment reduction obtained for the stochastic process remains valid in
both deterministic systems. Because the susceptible fraction in the SIR
model decreases continuously and no endemic fixed point with $i^*>0$
exists, the bifurcation analysis below focuses on the SIS model.

The behavior near the disease-free state is determined by the linear term
in $\Phi_q(i)$. Since $\mathsf{S}(q,1)=1$,
$\Phi_q(i)=M_1^{(\eta)}i+O(i^2)$. Linearization around $i=0$ gives
\begin{equation}
\frac{di}{dt}
=
\mu(\mathcal{R}_0-1)i+O(i^2),
\quad
\mathcal{R}_0
=
\frac{\beta M_1^{(\eta)}}{\mu}.
\label{eq:basic_reproduction_number}
\end{equation}
The disease-free state is stable for $\mathcal{R}_0<1$ and loses stability
at $\mathcal{R}_0=1$. In the SIR model, infection initially grows when
$s_0\mathcal{R}_0>1$, where $s_0$ is the initial susceptible fraction. A
linear term remains present even when $q>1$ because, at very low
prevalence, an exposed group typically contains only one infected
individual and $1^q=1$. Nonlinear effects become relevant only when several
infected individuals occur within the same group.

To separate the roles of the different moments, we introduce
$a_k=\mathsf{S}(q,k)M_k^{(\eta)}/M_1^{(\eta)}$, so that $a_1=1$ and
$a_k=0$ for $k>q$. In the rescaled time $\tau=\mu t$, the local expansion
of Eq.~\eqref{eq:deterministic_sis} is
\begin{equation}
\frac{di}{d\tau}
=
i\left[
(\mathcal{R}_0-1)
+
\mathcal{R}_0(a_2-1)i
+
\mathcal{R}_0(a_3-a_2)i^2
+
O(i^3)
\right].
\label{eq:general_local_normal_form}
\end{equation}
The parameter $\mathcal{R}_0$ depends only on $M_1^{(\eta)}$ and controls
the linear stability of the disease-free state. The coefficient
$a_2=\mathsf{S}(q,2)M_2^{(\eta)}/M_1^{(\eta)}$ determines the direction of
the nonlinear branch that meets $i=0$ at $\mathcal{R}_0=1$. For $a_2<1$,
the endemic branch emerges continuously. For $a_2>1$, the nonlinear
feedback is reinforcing and the branch becomes subcritical, allowing an
endemic state to persist for $\mathcal{R}_0<1$. At $a_2=1$, the quadratic
term vanishes. If $a_3-a_2<0$, this point marks the tricritical boundary
between continuous and discontinuous transitions.

For comparison, the linear kernel $q=1$ gives
$\Phi_1(i)=M_1^{(\eta)}i$ and the endemic fixed point
$i_{\mathrm{e}}^*=1-\mathcal{R}_0^{-1}$ for
$\mathcal{R}_0>1$. This branch grows continuously from $i=0$ as
$\mathcal{R}_0$ crosses one. For $\eta=1$, $M_1^{(1)}=1$, so
$\mathcal{R}_0=\beta/\mu$. For $\eta=0$,
$M_1^{(0)}=\langle n-1\rangle_Q$, and the invasion threshold therefore
depends on the mean number of partners in a group experienced by a
participant. Richer behavior appears for the quadratic kernel $q=2$. In this case,
$\Phi_2(i)=M_1^{(\eta)}i+M_2^{(\eta)}i^2$. Defining the nonlinear moment
ratio $\chi=M_2^{(\eta)}/M_1^{(\eta)}$ allows
Eq.~\eqref{eq:deterministic_sis} to be written as
\begin{equation}
\frac{di}{d\tau}
=
i\left[
\mathcal{R}_0(1-i)(1+\chi i)-1
\right].
\label{eq:q2_dimensionless_sis}
\end{equation}
Expanding this expression gives
$i'=i[(\mathcal{R}_0-1)+\mathcal{R}_0(\chi-1)i
-\mathcal{R}_0\chi i^2]$. Thus, $\mathcal{R}_0$ controls linear growth,
whereas $\chi$ sets the direction and strength of the nonlinear feedback.
For $\eta=0$,
$\chi=\langle(n-1)(n-2)\rangle_Q/\langle n-1\rangle_Q$, whereas for
$\eta=1$, $\chi=\langle n-2\rangle_Q$.

In addition to the disease-free state $i_0^*=0$, positive fixed points
satisfy $\mathcal{R}_0(1-i^*)(1+\chi i^*)=1$. For $\chi>0$, the two formal
roots are
\begin{equation}
i_{\pm}^*
=
\frac{
\chi-1
\pm
\sqrt{(1+\chi)^2-4\chi/\mathcal{R}_0}
}
{2\chi}.
\label{eq:q2_endemic_roots}
\end{equation}
When both roots lie in the physical interval $0<i^*<1$, the upper branch
$i_+^*$ is stable, whereas the lower branch $i_-^*$ is unstable and
separates the basins of the disease-free and endemic states. The case
$\chi=0$ reduces to linear incidence, with
$i_{\mathrm{e}}^*=1-\mathcal{R}_0^{-1}$. For $0\leq\chi<1$, the endemic branch emerges continuously as
$\mathcal{R}_0$ crosses one. At $\chi=1$, the quadratic term in the normal
form vanishes, and the positive fixed point satisfies
$i^*=\sqrt{1-\mathcal{R}_0^{-1}}$. Near the threshold,
$i^*\sim(\mathcal{R}_0-1)^{1/2}$, rather than the linear scaling obtained
for $\chi<1$. This change in scaling, together with the emergence of a
subcritical branch for $\chi>1$, identifies
$(\chi,\mathcal{R}_0)=(1,1)$ as a tricritical point. For $\chi>1$, a pair of positive fixed points is created through a
saddle-node bifurcation. The double-root condition gives
\begin{equation}
\mathcal{R}_{\mathrm{SN}}
=
\frac{4\chi}{(1+\chi)^2},
\quad
i_{\mathrm{SN}}
=
\frac{\chi-1}{2\chi}.
\label{eq:saddle_node_condition}
\end{equation}
Because $\mathcal{R}_{\mathrm{SN}}<1$, the bistable region is
$\mathcal{R}_{\mathrm{SN}}<\mathcal{R}_0<1$. Within this interval, the
disease-free and endemic states are both stable, while $i_-^*$ separates
their basins of attraction. The width of the hysteresis interval is
$\Delta\mathcal{R}=1-\mathcal{R}_{\mathrm{SN}}
=(\chi-1)^2/(1+\chi)^2$. Increasing $\chi$ does not shift the linear
stability boundary $\mathcal{R}_0=1$, but it lowers the threshold at which
the endemic branch first exists and broadens the bistable region.

\begin{figure*}[tb]
\centering
\includegraphics[width=\textwidth]
{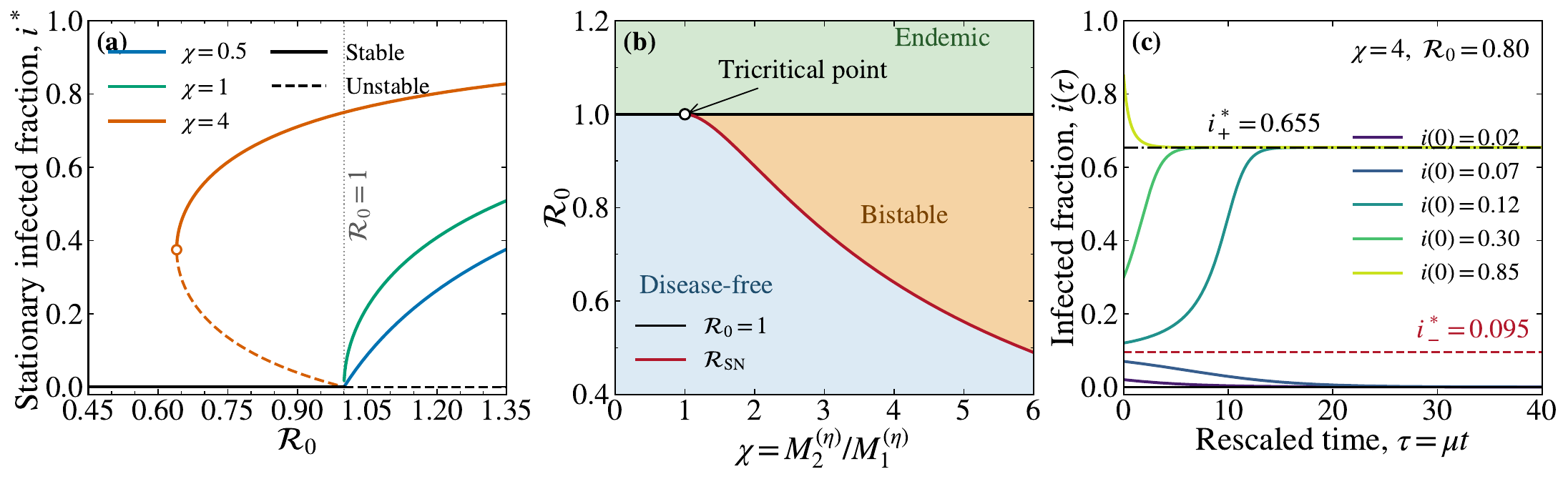}
\caption{Deterministic phase structure of the SIS model with the quadratic
infection kernel $q=2$. (a) Stationary infected fraction $i^*$ as a
function of $\mathcal{R}_0$ for $\chi=0.5$, $1$, and $4$. Solid and dashed
curves denote stable and unstable fixed points, respectively. The open
circle marks the saddle-node bifurcation for $\chi=4$, and the
vertical dotted line indicates $\mathcal{R}_0=1$. (b) Phase diagram in the
$(\chi,\mathcal{R}_0)$ plane. The black line marks the linear stability
boundary of the disease-free state, $\mathcal{R}_0=1$, while the red curve
shows the saddle-node boundary
$\mathcal{R}_{\mathrm{SN}}=4\chi/(1+\chi)^2$. The two boundaries meet at
the tricritical point $(\chi,\mathcal{R}_0)=(1,1)$. The blue, orange, and
green regions correspond to the disease-free, bistable, and
single-endemic-attractor regimes, respectively. (c) Evolution of the
infected fraction in the bistable regime for $\chi=4$ and
$\mathcal{R}_0=0.8$. The horizontal red dashed line marks the unstable
fixed point $i_-^*=0.095$, whereas the black dash-dotted line marks the
stable endemic fixed point $i_+^*=0.655$. Initial conditions below $i_-^*$
approach the disease-free state, whereas those above $i_-^*$ approach the
endemic state.}
\label{fig:phase_structure}
\end{figure*}

The fixed-point structure is summarized in
Fig.~\ref{fig:phase_structure}. Panels (a) and (b) are obtained analytically
from Eqs.~\eqref{eq:q2_endemic_roots} and
\eqref{eq:saddle_node_condition}, whereas panel (c) is obtained by
numerically integrating Eq.~\eqref{eq:q2_dimensionless_sis}. Representing
the results in the $(\chi,\mathcal{R}_0)$ plane separates the role of the
first moment in controlling linear invasion from that of the second moment
in controlling nonlinear feedback. Panel (a) shows how the nature of the transition changes as $\chi$
increases. For $\chi=0.5$, the endemic branch grows continuously from
$i^*=0$ at $\mathcal{R}_0=1$. The transition at $\chi=1$ remains
continuous but follows the square-root scaling associated with the
tricritical point. For $\chi=4$, a pair of positive fixed points appears
earlier through a saddle-node bifurcation at
$\mathcal{R}_{\mathrm{SN}}=0.64$ and $i_{\mathrm{SN}}=0.375$. By the time
$\mathcal{R}_0$ reaches one, the stable endemic branch has already reached
$i_+^*=0.75$. Increasing $\mathcal{R}_0$ through one can therefore produce
a jump from zero prevalence to a finite endemic level. Conversely, when
$\mathcal{R}_0$ is decreased along the endemic branch, the system returns
to the disease-free state only at $\mathcal{R}_{\mathrm{SN}}=0.64$. The
hysteresis interval consequently has width $\Delta\mathcal{R}=0.36$.

Panel (b) divides the parameter plane into three dynamical regimes. Below
the saddle-node curve, the disease-free state is the only attractor.
For $\chi>1$ and
$\mathcal{R}_{\mathrm{SN}}<\mathcal{R}_0<1$, the disease-free and endemic
states are both stable. For $\mathcal{R}_0>1$, the disease-free state is
unstable, leaving the endemic state as the only attractor in the physical
interval. The saddle-node curve and the line $\mathcal{R}_0=1$ meet
at $(\chi,\mathcal{R}_0)=(1,1)$, where the transition changes from
continuous to discontinuous. The decrease in
$\mathcal{R}_{\mathrm{SN}}$ with increasing $\chi$ shows that stronger
nonlinear amplification broadens the bistable region. Thus,
$\mathcal{R}_0<1$ guarantees local stability of the disease-free state but
does not necessarily exclude an endemic attractor. The dependence on the initial condition is shown in panel (c). For
$\chi=4$ and $\mathcal{R}_0=0.8$, the two positive fixed points are
$i_-^*\simeq0.095$ and $i_+^*\simeq0.655$. Trajectories initialized at
$i(0)=0.02$ and $0.07$ lie below $i_-^*$ and decay toward the disease-free
state. By contrast, the initial conditions $i(0)=0.12$, $0.30$, and $0.85$
lie above this boundary and all approach the same endemic state.
Trajectories starting from $0.12$ and $0.30$ increase toward $i_+^*$,
whereas the trajectory starting from $0.85$ decreases toward it. These
results confirm that $i_-^*$ is the boundary between the two basins of
attraction rather than merely an additional root of the fixed-point
equation. Epidemiologically, $i_-^*$ can be interpreted as a critical infection
mass. When $\mathcal{R}_0<1$, an initially rare infection declines because
the disease-free state is linearly stable. Once the initial prevalence
exceeds $i_-^*$, quadratic reinforcement becomes strong enough to balance
recovery and sustain the endemic branch. In panel (c), changing $i(0)$ from
$0.07$ to $0.12$ is sufficient to send the system to a different long-term
state. This example shows that $\mathcal{R}_0$ alone is not enough to assess
outbreak risk when transmission through groups is nonlinear. Discontinuous transitions, bistability, and critical infection masses have
also been reported in contagion models with higher-order interactions
\cite{iacopini2019,dearruda2020,matamalas2020,landry2020,stonge2021}.
The present reduction identifies the specific combinations of group-size
moments that control these phenomena when group sizes fluctuate between
exposure events.

Figure~\ref{fig:phase_structure} also separates the roles of the first two
moments. The moment $M_1^{(\eta)}$ sets the invasion threshold through
$\mathcal{R}_0$, whereas $M_2^{(\eta)}$ controls the bifurcation direction,
bistability, the critical infection mass, and the hysteresis width through
$\chi$. For a kernel of general order, the contribution of the second
moment is summarized by
$a_2=\mathsf{S}(q,2)M_2^{(\eta)}/M_1^{(\eta)}$. Two distributions may
therefore have the same invasion threshold because their
$M_1^{(\eta)}$ values coincide, yet exhibit different phase structures if
their $M_2^{(\eta)}$ values differ. The endemic states in Fig.~\ref{fig:phase_structure} are stable fixed points
of the deterministic equation. In a finite population, $I=0$ remains an
absorbing state, and stochastic fluctuations can ultimately drive the
system to extinction. The deterministic endemic branch therefore
corresponds to a long-lived quasistationary or metastable state in the
finite stochastic process~\cite{nasell1999,dickman2002}. The connection
between deterministic phase structure and exact stochastic equivalence is
examined in the next subsection using group-size distributions matched by
their factorial moments.

\subsection{Exact equivalence of group-size distributions}
\label{sec:moment_equivalence}

The moment-sufficiency theorem in Subsec.~\ref{sec:exact_reduction} states
that two group-size distributions can generate the same aggregate epidemic
process when the factorial moments accessed by the infection kernel
coincide. To test this result, we choose two participant-view distributions
with completely disjoint supports. The analysis in this subsection uses
$q=2$ and $\eta=0$, so the generator depends only on $M_1^{(0)}$ and
$M_2^{(0)}$. The choice $\eta=0$ allows the two distributions to be written
with simple rational probabilities. The same equivalence holds for
$\eta\neq0$ provided that the corresponding generalized moments are also
matched.

The first distribution contains only groups of sizes $3$ and $9$,
\begin{equation}
Q_A(n)
=
\frac{1}{2}\delta_{n,3}
+
\frac{1}{2}\delta_{n,9},
\label{eq:distribution_A}
\end{equation}
whereas the second contains groups of sizes $2$, $5$, and $12$,
\begin{equation}
Q_B(n)
=
\frac{1}{10}\delta_{n,2}
+
\frac{5}{7}\delta_{n,5}
+
\frac{13}{70}\delta_{n,12},
\label{eq:distribution_B}
\end{equation}
where $\delta_{n,m}$ denotes the Kronecker delta. Despite their different
forms and supports, the two distributions have
$\langle n\rangle_Q=6$, $\operatorname{Var}_Q(n)=9$,
$M_1^{(0)}=5$, and $M_2^{(0)}=29$. The agreement ends at third order:
$M_{3,A}^{(0)}=168$, whereas $M_{3,B}^{(0)}=201$. Their difference of
$33$ confirms that $Q_A$ and $Q_B$ are genuinely distinct distributions
rather than alternative representations of the same one. Because $\mathsf{S}(2,1)=\mathsf{S}(2,2)=1$, both distributions produce
the same mean infection hazard per susceptible individual,
\begin{equation}
\overline{\lambda}_A(I)
=
\overline{\lambda}_B(I)
=
\beta
\left[
5\frac{I}{N-1}
+
29\frac{I(I-1)}{(N-1)(N-2)}
\right].
\label{eq:matched_exact_infection_rate}
\end{equation}
The aggregate infection rate is $S\overline{\lambda}(I)$, whereas the
recovery rate remains $\mu I$. These rates coincide at every admissible
state $(S,I)$ and for every $N\geq12$. The result is therefore neither a
mean-field approximation nor a property that emerges only as
$N\rightarrow\infty$.

In the deterministic limit, the two distributions generate the same
exposure function $\Phi_2(i)=5i+29i^2$, reproduction number
$\mathcal{R}_0=5\beta/\mu$, and nonlinear moment ratio $\chi=29/5=5.8$.
The corresponding saddle-node threshold is
$\mathcal{R}_{\mathrm{SN}}\simeq0.50173$. The equivalence, however, is much
stronger than agreement between fixed points or phase diagrams. Because
the transition rates coincide, the two Markov generators are identical.
For the same initial condition,
\begin{equation}
\mathbb{P}_A\!\left[X(t)=x\right]
=
\mathbb{P}_B\!\left[X(t)=x\right]
\quad
\text{for every }t\text{ and }x,
\label{eq:equal_state_probability}
\end{equation}
where $X(t)=I(t)$ for SIS and $X(t)=(S(t),I(t))$ for SIR. This equality
covers the full state distribution and every statistic derived from it. The SIS predictions are obtained from the finite-state master equation on
$I=0,\ldots,N$. The mean prevalence is
$\langle i(t)\rangle=N^{-1}\sum_I I p_I(t)$, and the survival probability
is $\mathcal{S}(t)=1-p_0(t)$. For SIR, the final outbreak-size distribution
is calculated from absorption probabilities on the state space $(S,I)$.
As an independent check of the analytical reduction, the rates for $Q_A$
and $Q_B$ are evaluated separately from the microscopic sums over $n$ and
$\ell$, rather than by directly substituting
Eq.~\eqref{eq:matched_exact_infection_rate}. Details of the master-equation
solution, absorption-probability calculation, and Gillespie simulations are
given in Appendix~\ref{app:numerical_procedures}.

\begin{figure*}[t]
\centering
\includegraphics[width=0.8\textwidth]
{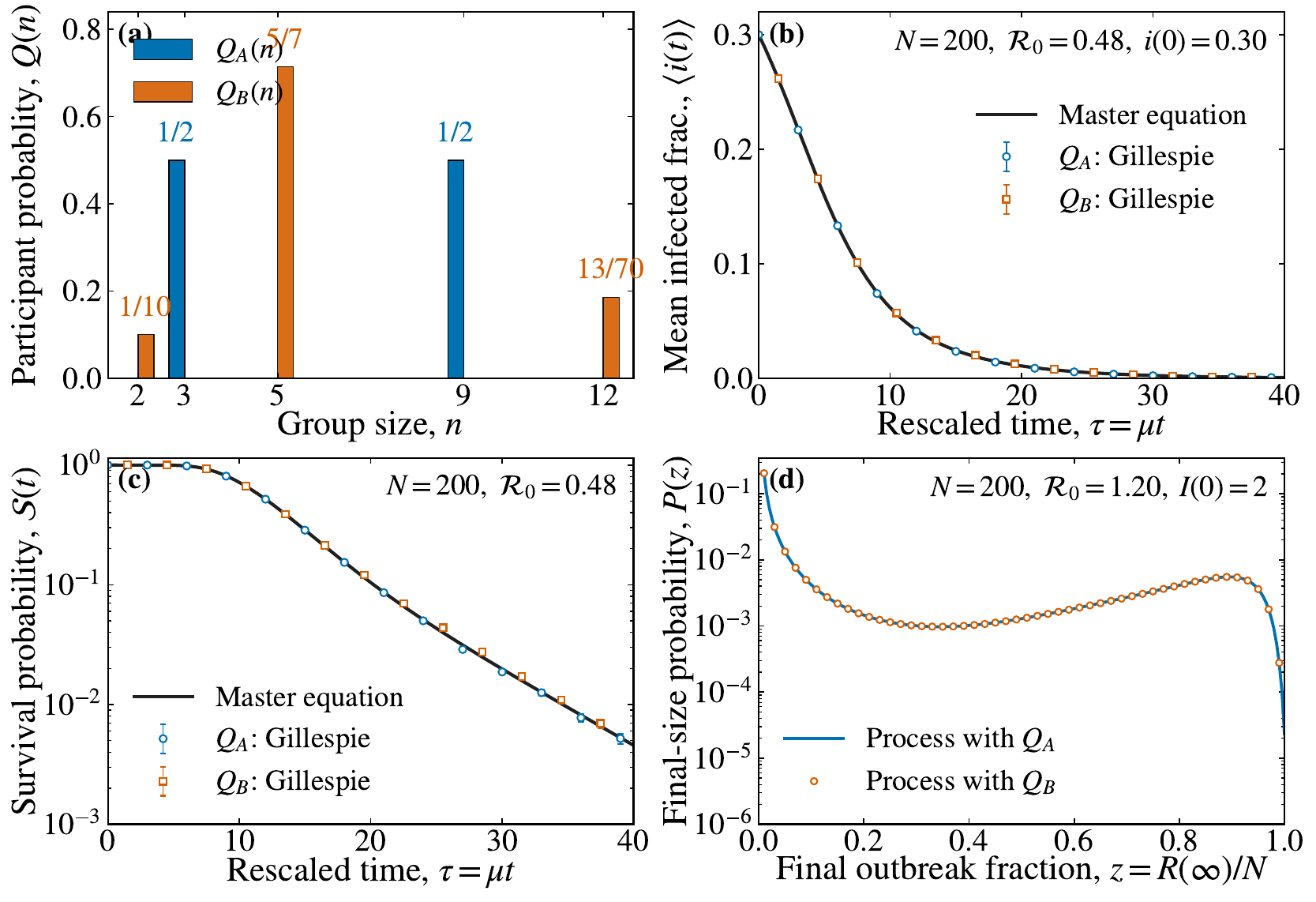}
\caption{Dynamical equivalence between two group-size distributions for
$q=2$ and $\eta=0$. (a) Participant-view distributions $Q_A(n)$ and
$Q_B(n)$ defined in Eqs.~\eqref{eq:distribution_A} and
\eqref{eq:distribution_B}. Although their supports are disjoint and
$M_{3,A}^{(0)}=168\neq201=M_{3,B}^{(0)}$, both distributions satisfy
$M_1^{(0)}=5$ and $M_2^{(0)}=29$. (b) Evolution of the mean SIS prevalence
for $N=200$, $\mathcal{R}_0=0.48$, and $i(0)=0.30$. (c) Survival
probability $\mathcal{S}(\tau)=\Pr(\mu T_{\mathrm{ext}}>\tau)$ for the same
SIS parameters. In panels (b) and (c), the black curves are solutions of
the master equation, whereas the blue circles and orange squares show
$N_{\mathrm{run}}=20\,000$ Gillespie simulations for $Q_A$ and $Q_B$,
respectively. Error bars indicate one standard error and are generally
smaller than the symbols. (d) Exact distribution of the final outbreak
fraction $z=R(\infty)/N$ in the SIR model for $N=200$,
$\mathcal{R}_0=1.20$, and $I(0)=2$; the blue curve and orange circles
represent $Q_A$ and $Q_B$, respectively. The overlap in panels (b)--(d)
shows that matching the first two factorial moments is sufficient to
produce identical laws for the aggregate epidemic process, despite the
different distribution shapes and third factorial moments.}
\label{fig:moment_equivalence}
\end{figure*}

Figure~\ref{fig:moment_equivalence}(a) makes clear that the observed
equivalence does not arise from any visual similarity between the
distributions: $Q_A$ has two support points with equal weights, whereas
$Q_B$ has three support points with unequal weights and a larger maximum
group size. Their third factorial moments differ by approximately $19.6\%$
relative to $M_{3,A}^{(0)}$, yet this difference is invisible to the
quadratic kernel because its transition rates depend only on the first two
factorial moments. Figure~\ref{fig:moment_equivalence}(b) shows the
corresponding SIS dynamics for $\mathcal{R}_0=0.48$, which lies slightly
below the saddle-node threshold
$\mathcal{R}_{\mathrm{SN}}\simeq0.50173$. The disease-free state is therefore
the only deterministic attractor, although the proximity to the threshold
produces a relatively slow decay. The mean prevalence decreases from its
initial value of $0.30$ to approximately $0.0616$ at $\tau=10$, $0.0107$
at $\tau=20$, and $5.72\times10^{-4}$ at $\tau=40$. Throughout this
transient, the master-equation solutions for $Q_A$ and $Q_B$ remain
indistinguishable, while the corresponding Gillespie estimates fluctuate
around the same curve within their statistical uncertainties.

Figure~\ref{fig:moment_equivalence}(c) examines the survival probability,
which contains information about extinction that cannot be recovered from
the mean prevalence alone. The survival probability is approximately
$0.717$ at $\tau=10$, decreases to $0.104$ at $\tau=20$, and reaches
$4.56\times10^{-3}$ at $\tau=40$. The condition
$\mathcal{S}(\tau)=1/2$ occurs at $\tau\simeq12.2$, giving the median
extinction time for the specified initial condition. The curves for $Q_A$
and $Q_B$ remain indistinguishable, confirming that equality of their
generators preserves the full extinction-time distribution rather than only
its mean. Figure~\ref{fig:moment_equivalence}(d) tests the equivalence using
the final outbreak size in the SIR model. For $\mathcal{R}_0=1.20$ and two
initially infected individuals, the distribution separates into small- and
large-outbreak regions. The probability mass at $z=0.01$ is approximately
$0.2045$ and corresponds to both initially infected individuals recovering
before causing any secondary infections, while a broader maximum near
$z\simeq0.895$ represents large outbreaks. Quantitatively, the probability
of a small outbreak with $z\leq0.10$ is approximately $0.583$, whereas the
probability of infecting at least half of the population,
$z\geq0.50$, is approximately $0.307$; the mean final outbreak fraction is
$\langle z\rangle\simeq0.285$. The agreement between $Q_A$ and $Q_B$ extends
over the entire distribution, from early extinction events to large
outbreaks.

The equality is also evident at the level of numerical precision. The
largest difference between the infection hazards evaluated separately for
the two distributions is $4.4\times10^{-16}$. For the SIS master equation,
the maximum difference between the state distributions is approximately
$2.1\times10^{-14}$, while the maximum difference between the SIR
final-size distributions is $5.6\times10^{-17}$. These values are at the
level of floating-point roundoff and support the generator equality derived
analytically. Equivalence here means equality in probability law, not equality between
individual sample paths. Two independent simulations using $Q_A$ and $Q_B$
need not produce the same sequence of events. Nevertheless, every aggregate
path has the same probability when the parameters and initial conditions
are identical. Group sizes in the simulations must also be sampled directly
from the participant-view distribution $Q(n)$. Sampling instead from the
event-level distribution $P(n)$ without size bias would define a different
stochastic process.

To verify that the dynamical collapse is not specific to this exposure
normalization or to a single population size, we repeat the exact
calculations for $\eta=1$ and $N=50,100,200,$ and $500$, as reported in
Appendix~\ref{app:eta_positive_multiN}. For every $N$, the prevalence
trajectory, SIS survival probability, and SIR final outbreak-size
distribution generated by the corresponding moment-matched pair coincide
to numerical precision. This calculation confirms that the equivalence
follows from the generalized factorial moments entering the Markov
generator, rather than from the special choice $\eta=0$, the
large-population limit, or the particular value $N=200$. At the same time,
the inequality $M_{3,A}^{(0)}\neq M_{3,B}^{(0)}$ provides a direct means of
probing the boundary of this equivalence: the two distributions cannot be
distinguished by a quadratic kernel, but become distinguishable once the
infection mechanism accesses the third factorial moment. The connection
between the kernel order and the first moment capable of separating two
distributions is examined in the next subsection.

\subsection{Moment hierarchy and structural nonidentifiability}
\label{sec:identifiability}

The exact reduction in Subsec.~\ref{sec:exact_reduction} also determines
how much information about the group-size distribution can be recovered
from aggregate epidemic dynamics. Using
$\ell^q=\sum_{k=1}^{q}\mathsf{S}(q,k)\ell_{\underline{k}}$, the
finite-population SIS transition rates can be written as
\begin{equation}
W_I^{+}
=
(N-I)\beta
\sum_{k=1}^{q}
\mathsf{S}(q,k)
M_k^{(\eta)}
\frac{I_{\underline{k}}}{(N-1)_{\underline{k}}},
\quad
W_I^{-}
=
\mu I,
\label{eq:finiteN_moment_hierarchy}
\end{equation}
where $\mathsf{S}(q,k)$ denotes a Stirling number of the second kind.
Equation~\eqref{eq:finiteN_moment_hierarchy} is exact for uniform sampling
without replacement. At fixed $q$ and $\eta$, the distribution $Q(n)$
enters the generator only through
$M_1^{(\eta)},\ldots,M_q^{(\eta)}$. Higher-order moments do not appear in
the transition rates and therefore cannot be learned from an epidemic
process governed by that kernel.

To express this hierarchy, we define the truncated moment vector
$\boldsymbol{m}_q[Q]=(M_1^{(\eta)},\ldots,M_q^{(\eta)})$. Two distributions
are equivalent at order $q$, written $Q_A\sim_q Q_B$, when
$M_{k,A}^{(\eta)}=M_{k,B}^{(\eta)}$ for all $1\leq k\leq q$. If $[Q]_q$
denotes the class of distributions equivalent to $Q$ at that order, then
$[Q]_{q+1}\subseteq[Q]_q$. Increasing the nonlinear order therefore
exposes one additional moment and can separate distributions that were
previously indistinguishable. For $\beta>0$, $1\leq q\leq N-1$, and the same values of
$N$, $\mu$, $q$, and $\eta$, matching the moments is not only sufficient
but also necessary for equality of the aggregate generators. The generator
equivalence class can therefore be characterized exactly by
\begin{equation}
\boldsymbol{m}_q[Q_A]
=
\boldsymbol{m}_q[Q_B]
\quad\Longleftrightarrow\quad
\mathcal{L}_A=\mathcal{L}_B.
\label{eq:generator_identifiability}
\end{equation}
The forward implication in
Eq.~\eqref{eq:generator_identifiability} follows directly from Theorem~1.
For the reverse implication, equality of the generators requires
$W_{I,A}^{+}=W_{I,B}^{+}$ for every state $I=1,\ldots,q$. At $I=1$, every
falling factorial with $k>1$ vanishes, so equality of the rates implies
$M_{1,A}^{(\eta)}=M_{1,B}^{(\eta)}$. Suppose that equality has already been
established up to order $m-1$. Evaluating the rates at $I=m$ leaves a new
contribution proportional to
$\beta\mathsf{S}(q,m)m!
[M_{m,A}^{(\eta)}-M_{m,B}^{(\eta)}]/
(N-1)_{\underline{m}}$. Its coefficient is nonzero, so
$M_{m,A}^{(\eta)}=M_{m,B}^{(\eta)}$. Repeating the argument through $m=q$
proves the reverse direction. The same reasoning applies to SIR by
selecting states with $S>0$.

Following the standard distinction between structural and practical
identifiability
\cite{bellman1970,miao2011,villaverde2019,wieland2021,heinrich2025},
three distinct sources of nonidentifiability must be separated in the
present setting. The first is generator nonidentifiability. If two
distributions belong to the same class $[Q]_q$, their generators and
complete aggregate path laws are identical. For the same initial condition
$x_0$, every time interval $[0,T]$, and every measurable set of paths
$\mathcal{B}$,
\begin{equation}
\mathbb{P}_A^{x_0}
\!\left[X_{[0,T]}\in\mathcal{B}\right]
=
\mathbb{P}_B^{x_0}
\!\left[X_{[0,T]}\in\mathcal{B}\right].
\end{equation}
This equality is intrinsic to the model: noise-free observations, perfect
temporal resolution, or access to all initial conditions cannot distinguish
$Q_A$ from $Q_B$ unless the microscopic protocol is changed. The same
statement applies to the SIR model after enlarging the state space to
$(S,I)$.

When $\beta$ is not known independently, the generator contains only the
combinations
$\vartheta_k=\beta\mathsf{S}(q,k)M_k^{(\eta)}$, with
$k=1,\ldots,q$. After time is rescaled by $\mu^{-1}$, these combinations
appear through $\mathcal{R}_0=\beta M_1^{(\eta)}/\mu$ and the ratios
$\mathsf{S}(q,k)M_k^{(\eta)}/M_1^{(\eta)}$ for $k\geq2$. Even perfect
reconstruction of the generator may therefore fail to separate the
transmission scale $\beta$ from the moments of the group-size distribution.
Moreover, the presence of a parameter in the generator does not guarantee
that it can be estimated from a single observable.

The second source is observable nonidentifiability. The identifiability of
epidemic models can depend strongly on the selected observation operator
and the type of available data~\cite{evans2005,dankwa2022}. Epidemic data usually
do not provide the full state distribution, but only selected projections,
such as mean prevalence, survival probability, or final outbreak size. If
$\boldsymbol{p}(\tau)$ denotes the state distribution and $\mathcal{H}$ is
an observation operator, the ideal output can be written as
$\boldsymbol{y}(\tau)=\mathcal{H}[\boldsymbol{p}(\tau)]$. For example,
mean prevalence is
$\langle i(\tau)\rangle=N^{-1}\sum_I I p_I(\tau)$, whereas survival
probability is $\mathcal{S}(\tau)=1-p_0(\tau)$. Equality of the generators
always implies
$\boldsymbol{p}_A(\tau)=\boldsymbol{p}_B(\tau)$ and
$\boldsymbol{y}_A(\tau)=\boldsymbol{y}_B(\tau)$, but the converse need not
hold. Two different generators may produce
$\boldsymbol{y}_A(\tau)=\boldsymbol{y}_B(\tau)$ exactly for a particular
observation operator and experimental design, even though other observables
or their path distributions differ. If $[Q]_{\mathcal{H}}$ denotes the class of distributions that produce the
same ideal output under the observation operator $\mathcal{H}$, then
$[Q]_q\subseteq[Q]_{\mathcal{H}}$, because generator equality necessarily
implies equality of every projected output. Observable-level degeneracy can
sometimes be reduced by combining several observables, initial conditions,
or experimental protocols \cite{ovchinnikov2022,bortner2024}. Such
additional information may distinguish different generators that happen to
produce the same output under a particular observation design, but it
cannot separate processes with identical generators. Exact equality under
$\mathcal{H}$ must also be distinguished from outputs that are merely close:
the former constitutes observable nonidentifiability, whereas a nonzero
difference that cannot be resolved against measurement noise or finite
sampling constitutes practical nonidentifiability.

The third source is practical nonidentifiability. Here, a moment affects
both the generator and the observable, but its influence is too weak
relative to measurement noise, sampling error, or correlations with other
parameters. The sensitivity of the infection rate to the $k$th moment is
\begin{equation}
\frac{\partial W_I^{+}}
     {\partial M_k^{(\eta)}}
=
(N-I)\beta\mathsf{S}(q,k)
\frac{I_{\underline{k}}}
     {(N-1)_{\underline{k}}},
\label{eq:finiteN_moment_sensitivity}
\end{equation}
whereas, in the large-$N$ limit, the sensitivity of the infected-fraction
drift behaves as
$\partial F(i)/\partial M_k^{(\eta)}
=\beta(1-i)\mathsf{S}(q,k)i^k+O(N^{-1})$. The factor $i^k$ shows that the
effect of higher-order moments decreases rapidly at low prevalence. A
moment may therefore be structurally accessible to a kernel with $q\geq k$
and yet remain difficult to estimate when the data cover only the early
phase of an outbreak.

Practical nonidentifiability can be examined through the sensitivity matrix
with elements
$J_{\alpha a}=\partial y(\tau_\alpha;\boldsymbol{\theta})/
\partial\theta_a$, where $\boldsymbol{\theta}$ contains the parameter
combinations to be estimated. For noise covariance
$\boldsymbol{\Sigma}$, the local information is summarized by
$\boldsymbol{\mathcal{I}}
=\boldsymbol{J}^{\mathsf{T}}\boldsymbol{\Sigma}^{-1}\boldsymbol{J}$.
Very small singular values of $\boldsymbol{J}$, or small eigenvalues of
$\boldsymbol{\mathcal{I}}$, indicate parameter directions that are
difficult to distinguish. Unlike generator degeneracy, this problem can be
mitigated by increasing the number of realizations, extending the
observation interval, using several initial conditions, or selecting a
protocol that is more sensitive to the moment of interest.

The present analysis does not estimate parameters from noisy data. The
sensitivity and information matrices above are introduced to explain why
higher-order moments may remain difficult to estimate even when they are
structurally present in the generator. The exact claim of this work is
therefore made at the generator level, whereas the discussion of practical
nonidentifiability concerns implications for inference design and is not
presented as a separate statistical-estimation result.

To demonstrate the moment hierarchy directly, we introduce the fixed
group-size distribution $Q_0(n)=\delta_{n,6}$ alongside $Q_A$ and $Q_B$
defined in Eqs.~\eqref{eq:distribution_A} and
\eqref{eq:distribution_B}. For $\eta=0$, their first three moment vectors
are
$\boldsymbol{m}_3[Q_0]=(5,20,60)$,
$\boldsymbol{m}_3[Q_A]=(5,29,168)$, and
$\boldsymbol{m}_3[Q_B]=(5,29,201)$. The pair $Q_0$ and $Q_A$ shares only
the first moment, whereas $Q_A$ and $Q_B$ match through second order. This
construction allows the influence of successive moments to be activated
step by step by increasing $q$.

\begin{figure*}[t]
\centering
\includegraphics[width=0.85\textwidth]
{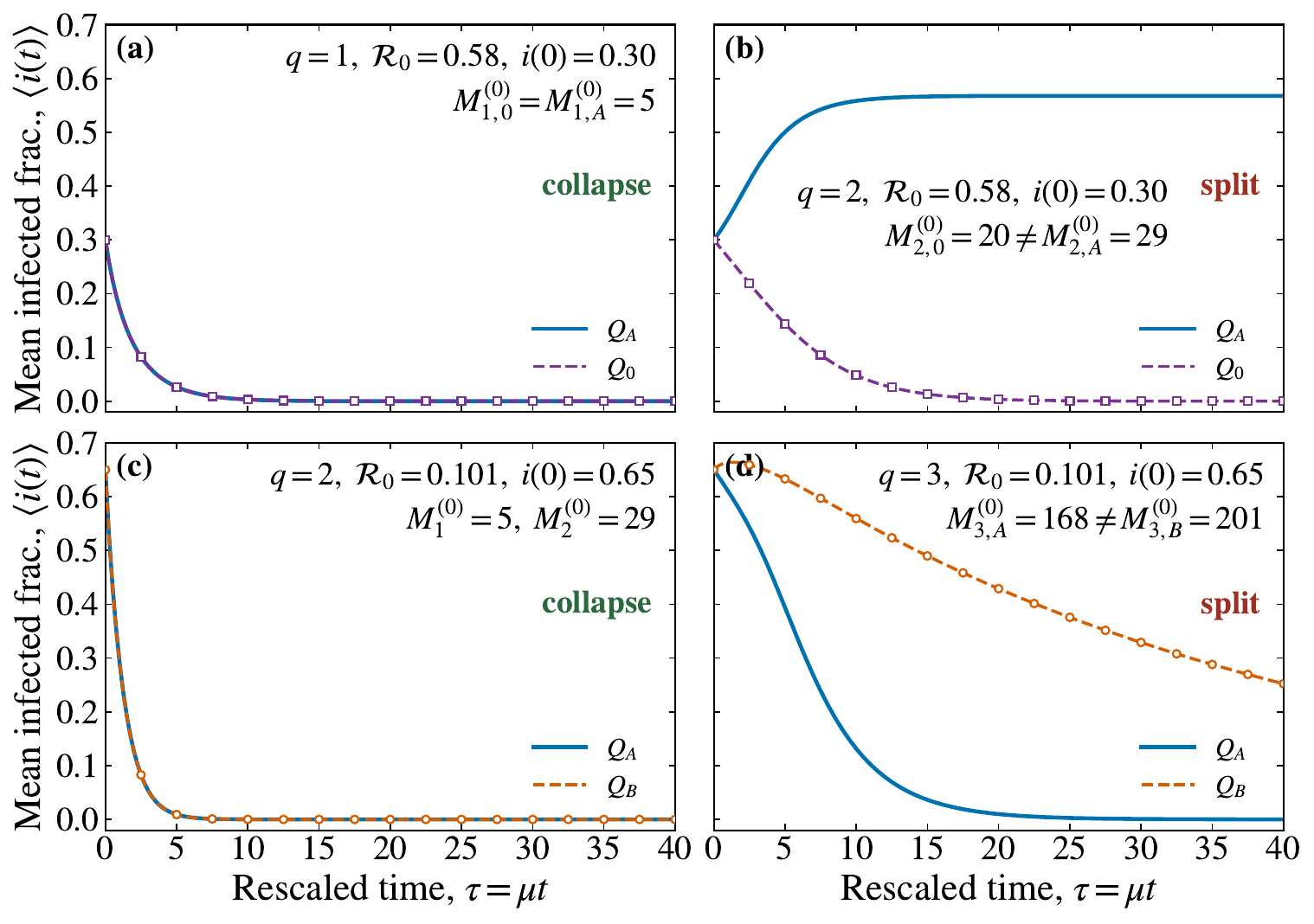}
\caption{Moment hierarchy and nonidentifiability in SIS dynamics. All curves
are obtained from numerical solutions of the finite-population master
equation for $N=200$, $\eta=0$, and rescaled time $\tau=\mu t$.
(a) The distributions $Q_0$ and $Q_A$ generate identical mean-prevalence
trajectories for $q=1$, $\mathcal{R}_0=0.58$, and $i(0)=0.30$ because
$M_{1,0}^{(0)}=M_{1,A}^{(0)}=5$. (b) For the same parameters with $q=2$,
the difference between $M_{2,0}^{(0)}=20$ and $M_{2,A}^{(0)}=29$
separates the curves. (c) The distributions $Q_A$ and $Q_B$ remain
dynamically identical for $q=2$, $\mathcal{R}_0=0.101$, and $i(0)=0.65$
because both have $M_1^{(0)}=5$ and $M_2^{(0)}=29$. (d) For the same
parameters with $q=3$, the curves separate because
$M_{3,A}^{(0)}=168$, whereas $M_{3,B}^{(0)}=201$. Solid and dashed curves
distinguish the distributions being compared. Open symbols serve only as
visual markers and do not represent separate simulation data.}
\label{fig:moment_hierarchy}
\end{figure*}

The dynamical consequences are shown in
Fig.~\ref{fig:moment_hierarchy}. Every curve is computed by first evaluating
the infection rate directly from the sums over group size and composition
and then solving the master equation separately for each distribution. The
reduced expression in Eq.~\eqref{eq:finiteN_moment_hierarchy} is not used
as numerical input. The collapse of the curves therefore provides an
independent test of the moment reduction. Panel (a) shows that the linear kernel accesses only $M_1^{(0)}$. Because
$Q_0$ and $Q_A$ both have $M_1^{(0)}=5$, the two generators are identical
even though the higher moments differ. At $\mathcal{R}_0=0.58<1$, the mean
prevalence falls from $0.30$ to approximately $3.06\times10^{-3}$ at
$\tau=10$ and $8.20\times10^{-9}$ at $\tau=40$. The maximum difference
between the curves is only $4.2\times10^{-16}$, so their overlap holds to
numerical precision. No improvement in data resolution under the $q=1$
protocol can reveal differences in the second or third moments because
neither moment enters the generator.

When the kernel order is increased to $q=2$, the difference between
$M_{2,0}^{(0)}$ and $M_{2,A}^{(0)}$ begins to affect the dynamics. For
$Q_0$, the nonlinear moment ratio is
$\chi_0=M_{2,0}^{(0)}/M_{1,0}^{(0)}=4$, giving
$\mathcal{R}_{\mathrm{SN},0}=0.640$, whereas for $Q_A$,
$\chi_A=29/5=5.8$ and
$\mathcal{R}_{\mathrm{SN},A}\simeq0.50173$. The value
$\mathcal{R}_0=0.58$ used in
Fig.~\ref{fig:moment_hierarchy}(b) lies between these thresholds, so the
system with $Q_0$ has only the disease-free attractor while that with $Q_A$
lies in the bistable regime. For $Q_A$, the unstable and stable positive
fixed points are $i_-^*\simeq0.198$ and $i_+^*\simeq0.629$, respectively.
Because the initial condition $i(0)=0.30$ lies above the basin boundary
$i_-^*$, the dynamics approaches the endemic branch, whereas the prevalence
under $Q_0$ continues to decay toward zero. At $\tau=40$, the mean prevalence
for $Q_0$ has fallen to approximately $1.06\times10^{-5}$, while that for
$Q_A$ remains near $0.568$. The latter value is lower than the deterministic
fixed point $i_+^*\simeq0.629$ because the unconditional master-equation
average includes both surviving trajectories near the metastable endemic
state and trajectories that have already been absorbed at $I=0$.

Figure~\ref{fig:moment_hierarchy}(c) tests the next pair in the hierarchy.
Although $Q_A$ and $Q_B$ have disjoint supports and different third
factorial moments, a quadratic kernel cannot access this difference. At
$\mathcal{R}_0=0.101$, the mean prevalence decreases from $0.65$ to
approximately $9.12\times10^{-3}$ at $\tau=5$ and
$1.03\times10^{-4}$ at $\tau=10$, while the maximum difference between the
two curves remains only $2.5\times10^{-16}$. Their collapse therefore
follows from equality of the generators, rather than from limited
observations or insufficient data. The difference becomes visible in
Fig.~\ref{fig:moment_hierarchy}(d), where increasing the kernel order to
$q=3$ activates $M_3^{(0)}$. At this order, the saddle-node thresholds are
$\mathcal{R}_{\mathrm{SN},A}\simeq0.1062$ and
$\mathcal{R}_{\mathrm{SN},B}\simeq0.0965$, and the value
$\mathcal{R}_0=0.101$ is deliberately chosen between them. The system
associated with $Q_A$ consequently has no positive endemic fixed point,
whereas the system associated with $Q_B$ has an unstable fixed point
$i_-^*\simeq0.503$ and a stable endemic fixed point
$i_+^*\simeq0.696$. Since the initial condition satisfies
$i(0)=0.65>i_-^*$, the deterministic dynamics for $Q_B$ approaches the
endemic state, while that for $Q_A$ approaches the disease-free state.

In the finite population, this distinction appears as a difference in
persistence times. The separation between the mean prevalences reaches
approximately $0.456$ at $\tau\simeq13.75$. At $\tau=40$, the prevalence
for $Q_A$ has fallen to approximately $5.32\times10^{-5}$, whereas that
for $Q_B$ remains near $0.252$. The $Q_B$ curve continues to decline even
though the deterministic equation has a stable fixed point near $0.696$,
because the unconditional mean also includes paths that have reached the
absorbing state $I=0$. The deterministic endemic branch appears as a
long-lived metastable state in the finite population.

The collapse in Figs.~\ref{fig:moment_hierarchy}(a) and
\ref{fig:moment_hierarchy}(c) does not depend on population size: $N$
controls the magnitude of stochastic fluctuations and the lifetime of
metastable states, but does not affect generator equality provided that all
group sizes are admissible. By contrast, the separation in
Figs.~\ref{fig:moment_hierarchy}(b) and
\ref{fig:moment_hierarchy}(d) occurs because increasing the kernel order
exposes the first unmatched factorial moment. This alternating pattern of
collapse and separation directly illustrates the nesting relation
$[Q]_{q+1}\subseteq[Q]_q$. More specifically,
Fig.~\ref{fig:moment_hierarchy} demonstrates nonidentifiability at the
generator level: the curves in panels (a) and (c) coincide because every
aggregate transition rate is identical, a statement stronger than agreement
of the mean prevalence alone. Observable nonidentifiability would instead
arise if different generators produced the same ideal output under a
particular observation operator, whereas practical nonidentifiability
occurs when the outputs differ but their separation is too small to resolve
in the presence of noise or finite data. The collapse in panels (a) and (c)
should therefore not be attributed to limited data resolution, while the
separation in panels (b) and (d) shows that changing the transmission
protocol can reveal information about the next factorial moment.

The appropriate remedy depends on the source of nonidentifiability.
Generator degeneracy requires a change in protocol, such as increasing
$q$, changing $\eta$, or adding a transmission channel involving the next
moment. Observable degeneracy can be reduced by combining mean prevalence
with survival probability, final-size distributions, and multiple initial
conditions. Practical nonidentifiability instead calls for more
realizations, a broader prevalence range, and an analysis of parameter
uncertainty. A single aggregate epidemic time series generated under one
protocol is therefore generally insufficient to determine the full form of
$Q(n)$ uniquely.

\subsection{Limits of moment equivalence}
\label{sec:breakdown}

The moment equivalence derived above is exact as long as the two processes
follow the same microscopic rules. These rules include annealed
group formation, uniform sampling without replacement, a distribution
$Q(n)$ that does not change with the epidemiological state, and a
fixed-order polynomial transmission kernel. Two issues delimit the result.
First, knowledge of finitely many moments need not determine $Q(n)$
uniquely. Second, changing the transmission mechanism can expose moments
that did not previously affect the dynamics.

To examine the first issue, suppose that the support is known and contains
$m$ group sizes, $\mathcal{G}=\{n_1,\ldots,n_m\}$. Their probabilities form
the vector
$\boldsymbol{Q}=(Q(n_1),\ldots,Q(n_m))^{\mathsf{T}}$, while normalization
and the moments through order $q$ define the linear system
$\boldsymbol{b}_q=\boldsymbol{A}_q^{(\eta)}\boldsymbol{Q}$. The first row
of $\boldsymbol{A}_q^{(\eta)}$ consists of ones and enforces normalization,
whereas the row associated with moment order $k$ has entries
$(n_j-1)_{\underline{k}}/(n_j-1)^\eta$. If
$\operatorname{rank}(\boldsymbol{A}_q^{(\eta)})=m$, these constraints
uniquely determine the distribution on the specified support; if the rank
is smaller than $m$, nonzero perturbations of $\boldsymbol{Q}$ exist that
preserve normalization and every measured moment. An interior distribution
$\boldsymbol{Q}_0$ may therefore belong to the continuous family
$\boldsymbol{Q}_{\theta}
=\boldsymbol{Q}_0+\theta\boldsymbol{v}$, where
$\boldsymbol{v}\in\ker(\boldsymbol{A}_q^{(\eta)})$, and every member of
this family generates the same aggregate epidemic generator as long as all
probabilities remain nonnegative. The local dimension of the corresponding
equivalence class is
$m-\operatorname{rank}(\boldsymbol{A}_q^{(\eta)})$. Since the rank cannot
exceed $q+1$, a support containing $m>q+1$ group sizes necessarily leaves
at least $m-q-1$ directions unidentifiable. Structural
nonidentifiability is therefore not confined to isolated pairs of
distributions, but can extend to continuous families within the probability
simplex.

Formally, a distribution on finite support can be reconstructed if
sufficiently many independent moments are available. Reconstruction from
high-order moments, however, is usually sensitive to measurement error. A
further difficulty arises when $\beta$ is not known independently, because
the generator contains only the combinations $\beta M_k^{(\eta)}$. After
time is normalized by $\mu^{-1}$, this information appears through
$\mathcal{R}_0$ and the ratios
$M_k^{(\eta)}/M_1^{(\eta)}$. Full rank of the moment map therefore does not
necessarily separate the influence of the group-size distribution from the
baseline transmission scale.

The mechanistic boundary can be tested by adding a transmission channel one
order higher,
\begin{equation}
\lambda_{n,\ell}^{(q,\varepsilon)}
=
\frac{\beta}{(n-1)^\eta}
\left[
\ell^q+\varepsilon\ell_{\underline{q+1}}
\right],
\quad
\varepsilon\geq0.
\label{eq:perturbed_infection_kernel}
\end{equation}
The parameter $\varepsilon$ controls the strength of the additional
channel. At $\varepsilon=0$, the original kernel is recovered. Using
$\ell_{\underline{q+1}}$ ensures that the new channel activates only
$M_{q+1}^{(\eta)}$ without changing the coefficients of lower-order
moments. The aggregate infection rate consequently gains the term
$(N-I)\beta\varepsilon M_{q+1}^{(\eta)}
I_{\underline{q+1}}/(N-1)_{\underline{q+1}}$.

If $Q_A\sim_q Q_B$ but
$M_{q+1,A}^{(\eta)}\neq M_{q+1,B}^{(\eta)}$, the difference between their
infection rates is
\begin{equation}
W_{I,B}^{+}-W_{I,A}^{+}
=
(N-I)\beta\varepsilon
\left[
M_{q+1,B}^{(\eta)}-M_{q+1,A}^{(\eta)}
\right]
\frac{I_{\underline{q+1}}}
     {(N-1)_{\underline{q+1}}}.
\label{eq:equivalence_breakdown_rate}
\end{equation}
For the pair $Q_A$ and $Q_B$ considered here,
$M_{3,B}^{(0)}-M_{3,A}^{(0)}=33$. With the quadratic kernel, the difference
becomes
$33(N-I)\beta\varepsilon I_{\underline{3}}/
(N-1)_{\underline{3}}$. This correction vanishes for $I<3$ and is cubic
near the disease-free state. The additional channel therefore leaves the
linear invasion threshold unchanged but can become important at
intermediate and high prevalence. For $q=2$ and $\eta=0$, the deterministic limit for the process using
$Q_X$, with $X\in\{A,B\}$, is
\begin{equation}
\frac{di}{d\tau}
=
i\left\{
\mathcal{R}_0(1-i)
\left(1+\chi i+\zeta_X i^2\right)-1
\right\},
\label{eq:perturbed_q2_deterministic_sis}
\end{equation}
where $\zeta_X
=
\varepsilon M_{3,X}^{(0)}/M_1^{(0)}$. Here, $\tau=\mu t$,
$\mathcal{R}_0=\beta M_1^{(0)}/\mu$, and
$\chi=M_2^{(0)}/M_1^{(0)}$. The two distributions retain the same
$\mathcal{R}_0$ and $\chi$ but acquire different values of $\zeta_X$.
Their different third moments can therefore change the phase structure
without altering the linear invasion threshold.

The saddle-node threshold follows from
$\mathcal{R}_{\mathrm{SN},X}H_X(i_{\mathrm{SN},X})=1$ and
$H_X'(i_{\mathrm{SN},X})=0$, where
$H_X(i)=(1-i)(1+\chi i+\zeta_Xi^2)$. The physical solution is
\begin{equation}
\begin{aligned}
i_{\mathrm{SN},X}
&=
\frac{
\zeta_X-\chi+
\sqrt{(\zeta_X-\chi)^2+3\zeta_X(\chi-1)}
}
{3\zeta_X}, \\
\mathcal{R}_{\mathrm{SN},X}
&=
\frac{1}{
(1-i_{\mathrm{SN},X})
(1+\chi i_{\mathrm{SN},X}+\zeta_Xi_{\mathrm{SN},X}^{2})
}.
\label{eq:perturbed_saddle_node_threshold}
\end{aligned}
\end{equation}
As $\zeta_X\rightarrow0$, these expressions reduce to
$i_{\mathrm{SN}}=(\chi-1)/(2\chi)$ and
$\mathcal{R}_{\mathrm{SN}}=4\chi/(1+\chi)^2$, recovering the threshold of
the unperturbed quadratic kernel.

For a prescribed $\mathcal{R}_0$, the critical value
$\varepsilon_{c,X}$ can be obtained without scanning $\varepsilon$
numerically. The critical prevalence $i_c$ is the physical root of
$\mathcal{R}_0(1-i_c)^2(2+\chi i_c)=2-3i_c$, and the perturbation strength
is
\begin{equation}
\varepsilon_{c,X}
=
\frac{M_1^{(0)}}{M_{3,X}^{(0)}}
\frac{
2\chi i_c-(\chi-1)
}{
i_c(2-3i_c)
}.
\label{eq:critical_perturbation_strength}
\end{equation}
This expression shows that a distribution with larger $M_3^{(0)}$ reaches
bistability under a weaker perturbation. For
$M_1^{(0)}=5$, $M_2^{(0)}=29$, and
$\mathcal{R}_0=0.48$, we obtain $i_c\simeq0.43636$ and
$\zeta_c\simeq0.86826$. The critical strengths are
$\varepsilon_{c,A}\simeq0.02584$ and
$\varepsilon_{c,B}\simeq0.02160$. The ratio
$\varepsilon_{c,B}/\varepsilon_{c,A}=168/201$ shows that the difference
between the two thresholds follows directly from the third moments.

\begin{figure*}[t]
\centering
\includegraphics[width=\textwidth]
{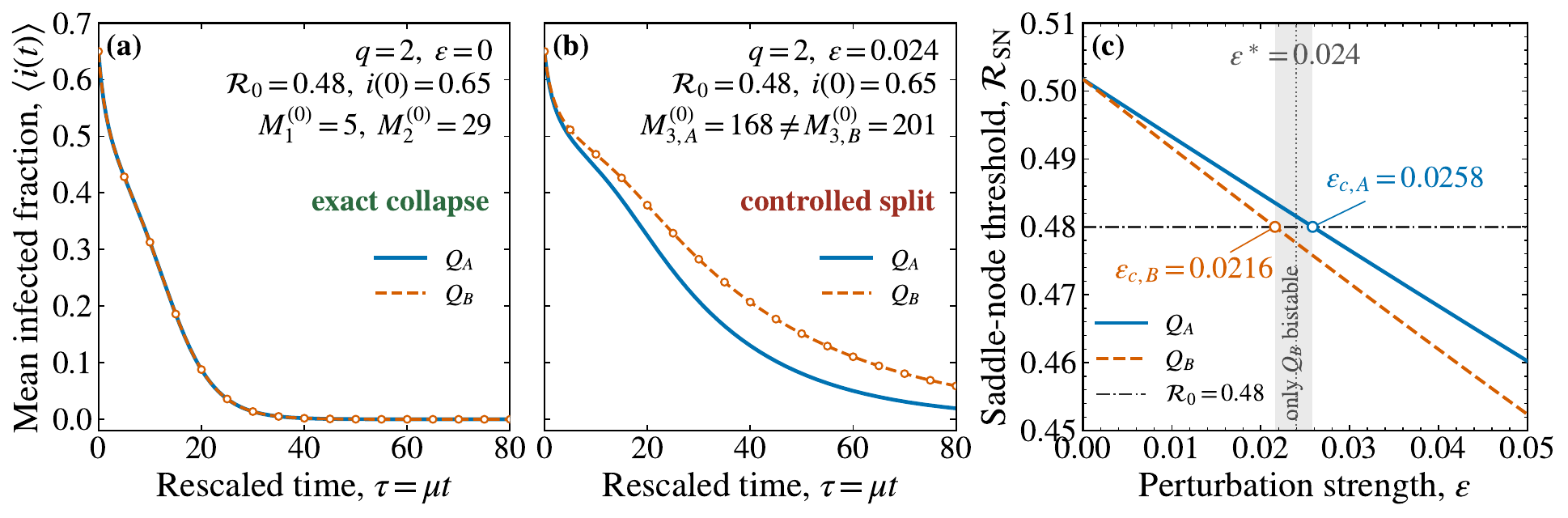}
\caption{Breakdown of moment equivalence after activation of a higher-order
transmission channel. Panels (a) and (b) show the mean SIS prevalence
obtained from the master equation for $N=500$, $q=2$, $\eta=0$,
$\mathcal{R}_0=0.48$, and $i(0)=0.65$. (a) At $\varepsilon=0$, the results
for $Q_A$ and $Q_B$ coincide because both distributions have
$M_1^{(0)}=5$ and $M_2^{(0)}=29$. (b) At $\varepsilon=0.024$, the channel
$\varepsilon\ell_{\underline{3}}$ activates the difference between
$M_{3,A}^{(0)}=168$ and $M_{3,B}^{(0)}=201$, causing the curves to
separate. (c) Saddle-node threshold
$\mathcal{R}_{\mathrm{SN}}$ as a function of $\varepsilon$. The horizontal
dash-dotted line marks $\mathcal{R}_0=0.48$, while the open circles indicate
$\varepsilon_{c,B}=0.0216$ and $\varepsilon_{c,A}=0.0258$. The shaded
region is the perturbation interval in which only the $Q_B$ system is
bistable. The vertical dotted line marks $\varepsilon^\ast=0.024$, as used
in panel (b). The solid blue and dashed orange curves represent $Q_A$ and
$Q_B$, respectively. Symbols in panels (a) and (b) are visual markers and
do not represent separate simulation data.}
\label{fig:equivalence_breakdown}
\end{figure*}

The analytical predictions are tested using the finite-population master
equation for $N=500$ and the initial condition $i(0)=0.65$, with the
infection rates evaluated directly from the sums over group size and
composition rather than from the reduced moment expression. This procedure
provides an independent check of the predicted breakdown of moment
equivalence, and the results are presented in
Fig.~\ref{fig:equivalence_breakdown}. In panel (a), the two curves coincide
throughout the evolution because the unperturbed kernel depends only on
$M_1^{(0)}$ and $M_2^{(0)}$. The mean prevalence decreases from $0.65$ to
approximately $0.313$ at $\tau=10$, $0.0879$ at $\tau=20$, and
$1.89\times10^{-3}$ at $\tau=40$, while the maximum difference between the
two curves is only about $1.1\times10^{-14}$ and can be attributed to
floating-point roundoff. Once the perturbation is set to
$\varepsilon=0.024$, the larger value
$M_{3,B}^{(0)}>M_{3,A}^{(0)}$ gives the $Q_B$ system a higher infection
rate. As shown in panel (b), the prevalences for $Q_A$ and $Q_B$ are
approximately $0.324$ and $0.378$, respectively, at $\tau=20$, and their
difference reaches a maximum of about $0.0775$ near $\tau\simeq37$. By
$\tau=80$, the prevalence has decreased to approximately $0.0194$ for
$Q_A$ but remains near $0.0589$ for $Q_B$. This separation is consistent
with the additional deterministic contribution proportional to
$(1-i)i^3$.

Figure~\ref{fig:equivalence_breakdown}(c) explains how this weak
higher-order channel produces a measurable separation. At
$\varepsilon=0$, both systems have
$\mathcal{R}_{\mathrm{SN}}\simeq0.50173$. Increasing $\varepsilon$ lowers
the saddle-node threshold for both distributions, but the decrease is
faster for $Q_B$ because its third factorial moment is larger. At
$\mathcal{R}_0=0.48$, the $Q_B$ system enters the bistable regime at
$\varepsilon_{c,B}\simeq0.02160$, whereas $Q_A$ does so only at
$\varepsilon_{c,A}\simeq0.02584$. The value
$\varepsilon^\ast=0.024$ used in panel (b) lies between these critical
values and gives
$\mathcal{R}_{\mathrm{SN},A}\simeq0.48153$ and
$\mathcal{R}_{\mathrm{SN},B}\simeq0.47762$. Thus, at
$\mathcal{R}_0=0.48$, the $Q_A$ system still has only the disease-free
attractor, while $Q_B$ already possesses a pair of endemic fixed points.
Because $i(0)=0.65$ lies above the basin boundary
$i_-^*\simeq0.397$, the deterministic $Q_B$ dynamics approaches the stable
endemic state at $i_+^*\simeq0.480$. The $Q_B$ curve in panel (b)
nevertheless continues to decrease because it represents the unconditional
mean of a finite population: $I=0$ remains absorbing, and trajectories that
have already become extinct continue to lower the ensemble average. The
deterministic endemic branch therefore appears in the stochastic process as
a metastable state with a finite lifetime, while proximity to the
saddle-node threshold makes fluctuations more effective at driving
trajectories out of that state.

Equivalence may also be broken by other changes to the microscopic
mechanism. Nonpolynomial kernels, including threshold functions and
saturating responses, generally probe a larger set of moments, while a
group-size-dependent transmission scale, $\beta=\beta(n)$, replaces the
relevant coefficients by
$\sum_nQ(n)\beta(n)(n-1)_{\underline{k}}/(n-1)^\eta$. Likewise, changing
$\eta$ can distinguish distributions whose generalized moments were matched
under only one exposure normalization, so equality under one transmission
protocol need not persist when that protocol is modified. The annealed mixing assumption is equally important. In quenched
higher-order structures, repeated group membership, hyperedge overlap, and
cross-order degree correlations introduce dynamical information that is not
contained in the annealed group-size distribution alone
\cite{burgio2024,kim2024,malizia2025overlap,
malizia2025heterogeneity,guzman2026}. Group composition is then no longer
described by a hypergeometric distribution determined solely by $I$, and
the infection rates depend on correlations among epidemiological states,
hyperedge sizes, and individual group memberships. Closure at the level of
$Q(n)$ is likewise lost if participation changes during an outbreak or
becomes correlated with epidemiological state. By themselves, however,
finite population size and the presence of an absorbing state do not break
the equivalence, provided that the two processes obey the same microscopic
rules. These features alter the observed dynamics, including fluctuations
and extinction statistics, but do not affect equality of the Markov
generators. The resulting equivalence classes are therefore specific to the
transmission protocol, particularly the choices of $q$ and $\eta$, the
sampling rule, and the mixing pattern. Combining observations from several
protocols that access different moment orders may consequently reduce the
set of group-size distributions consistent with the data.

\section{Conclusion}
\label{sec:conclusion}

We have studied SIS and SIR dynamics in a well-mixed population where
transmission occurs through temporary groups of random size. Exact averaging
over group composition shows that, for a polynomial infection kernel of order
$q$, the finite-population Markov generator depends on the group-size
distribution $Q(n)$ only through the generalized factorial moments
$M_1^{(\eta)},\ldots,M_q^{(\eta)}$. Matching these moments through order $q$
therefore implies more than identical epidemic thresholds or deterministic
limits: it makes the entire aggregate stochastic processes identical in law.
This equivalence covers transient dynamics, fluctuations, SIS extinction-time
distributions, and SIR final outbreak-size distributions.

The deterministic limit assigns distinct roles to successive levels of the
moment hierarchy. The first moment determines the linear stability of the
disease-free state through $\mathcal{R}_0$, whereas the second moment controls
nonlinear feedback and the resulting bifurcation structure. For a quadratic
kernel, the ratio
$\chi=M_2^{(\eta)}/M_1^{(\eta)}$ separates continuous transitions, the
tricritical point, and a bistable regime with hysteresis. The pair
$Q_A(n)$ and $Q_B(n)$ provides a direct illustration: the two distributions
have disjoint supports and different third moments, yet generate identical
SIS and SIR dynamics for $q=2$ because their first two moments coincide.
When the kernel order is raised to $q=3$, the third moment enters the
transition rates and the two dynamics become distinguishable.

These results also establish a limit on what can be inferred from aggregate
epidemic data. Generator-level nonidentifiability arises when two
distributions share the same vector of relevant moments; this degeneracy
cannot be removed by increasing the temporal resolution or extending the
observation period under the same transmission protocol. Observable-level
nonidentifiability may occur even when the generators differ if the measured
statistics do not resolve that difference, whereas practical
nonidentifiability results from noise, limited data, or weak sensitivity.
The controlled perturbation through the channel
$\varepsilon\ell_{\underline{q+1}}$ shows that activating a single additional
moment can shift the saddle-node threshold and place two previously
equivalent distributions in different dynamical regimes.

The equivalence derived here rests on the microscopic protocol: annealed
mixing, uniform sampling without replacement, a group-size distribution
$Q(n)$ that is independent of the epidemic state, and a polynomial kernel of
fixed order. Topological correlations in a quenched hypergraph,
state-dependent participation, or nonpolynomial infection kernels generally
expose information beyond the truncated moment vector. Extending the analysis
to these settings would clarify how far the equivalence classes persist in
more realistic contact structures. From an inference perspective, the results
suggest combining several exposure protocols with different nonlinear orders
or normalization rules, thereby accessing additional moments and narrowing
the set of group-size distributions that remain observationally
indistinguishable.
\section*{Acknowledgments}
\textbf{Roni~Muslim} was supported by the YST Program of the Asia Pacific Center for Theoretical Physics (APCTP), funded by the Science and Technology Promotion Fund and Lottery Fund of the Korean Government, and by the Management Talent Program of the National Research and Innovation Agency of Indonesia (BRIN).

\section*{AUTHOR DECLARATIONS}
\subsection*{Conflict of Interest}
The author has no conflicts to disclose.

\subsection*{Author Contributions}
\textbf{Roni Muslim:} Conceptualization; Formal analysis; Investigation;
Methodology; Software; Validation; Visualization; Writing--original draft;
Writing--review and editing.

\appendix
\section{Numerical procedures}
\label{app:numerical_procedures}

All calculations were performed in double precision, and the results are
reported in terms of the rescaled time $\tau=\mu t$. The recovery rate was
set to $\mu=1$, while $\beta$ was determined from
$\beta=\mathcal{R}_0\mu/M_1^{(\eta)}$ so that $\mathcal{R}_0$ could be
compared directly across distributions. The deterministic curves in
Fig.~\ref{fig:phase_structure} were obtained from analytical fixed-point
solutions whenever available, while the time trajectories were computed
using an adaptive Runge--Kutta integrator with relative and absolute
tolerances of $10^{-10}$ and $10^{-12}$, respectively. For the
finite-population stochastic calculations, the infection rate associated
with each $Q_X(n)$, where $X\in\{0,A,B\}$, was evaluated directly from the
microscopic sum
$\overline{\lambda}_X(I)=\sum_nQ_X(n)\sum_\ell
\Pr(\ell\mid n,I)\lambda_{n,\ell}$ rather than from its reduced moment
representation. This provides an independent numerical check of
Eq.~\eqref{eq:averaged_infection_rate} and Theorem~1. For the SIS model, the
initial condition was set to $p_I(0)=\delta_{I,I_0}$, and the state
distribution was propagated through the master equation
\begin{equation}
\frac{dp_I(t)}{dt}
=
W_{I-1}^{+}p_{I-1}(t)
+
W_{I+1}^{-}p_{I+1}(t)
-
\left(W_I^{+}+W_I^{-}\right)p_I(t),
\label{eq:sis_master_equation}
\end{equation}
where $W_I^{+}=(N-I)\overline{\lambda}(I)$,
$W_I^{-}=\mu I$, and terms outside the state space were set to zero. This
equation was integrated using the second-order Crank--Nicolson scheme with
an internal time step $\Delta\tau=10^{-3}$. The linear system at each step
has a tridiagonal structure and was solved by tridiagonal LU factorization.

The mean prevalence and survival probability were calculated as
$\langle i(\tau)\rangle=N^{-1}\sum_I I p_I(\tau)$ and
$\mathcal{S}(\tau)=1-p_0(\tau)=\Pr(\mu T_{\mathrm{ext}}>\tau)$,
respectively. For the SIR model, the system state was represented by $(S,I)$,
with infection rate
$W_{S,I}^{\mathrm{inf}}=S\overline{\lambda}(I)$ and recovery rate
$W_{S,I}^{\mathrm{rec}}=\mu I$. Because the infection transition
$(S,I)\to(S-1,I+1)$ and the recovery transition
$(S,I)\to(S,I-1)$ drive the process toward the absorbing set $I=0$ without
forming cycles, the final outbreak-size distribution was calculated from
the recursion
\begin{equation}
F_{S,I}(S_f)
=
p_{\mathrm{inf}}(S,I)F_{S-1,I+1}(S_f)
+
p_{\mathrm{rec}}(S,I)F_{S,I-1}(S_f),
\label{eq:sir_absorption_recursion}
\end{equation}
with $F_{S,0}(S_f)=\delta_{S,S_f}$,
$p_{\mathrm{inf}}=W_{S,I}^{\mathrm{inf}}/
(W_{S,I}^{\mathrm{inf}}+W_{S,I}^{\mathrm{rec}})$, and
$p_{\mathrm{rec}}=1-p_{\mathrm{inf}}$. For $R(0)=0$, the final outbreak
fraction is $z=(N-S_f)/N$. The Gillespie simulations used the total rate
$W_{\mathrm{tot}}=W^{\mathrm{inf}}+W^{\mathrm{rec}}$, the interevent time
$\Delta t=-\ln u_1/W_{\mathrm{tot}}$, and an infection event was selected
whenever $u_2<W^{\mathrm{inf}}/W_{\mathrm{tot}}$, where $u_1$ and $u_2$
are independent uniform random numbers on $(0,1)$. Each SIS trajectory was
terminated upon reaching $I=0$ or the prescribed maximum observation time,
whereas each SIR trajectory was terminated when $I=0$. The Gillespie symbols
in Figs.~\ref{fig:moment_equivalence}(b) and
\ref{fig:moment_equivalence}(c) were obtained from
$N_{\mathrm{run}}=20000$ independent realizations for each distribution
and initial condition. The symbols represent ensemble averages, and the
error bars denote one standard error. In every calculation, we checked the
normalization and nonnegativity of the probability distribution, as well as
the absorbing character of the state $I=0$. Rates for moment-matched
distributions were always evaluated separately, ensuring that the overlap
of the curves follows from equality of the microscopic transition rates
rather than from assigning the same reduced generator to both processes.

\section{Equivalence under normalized exposure and at finite population sizes}
\label{app:eta_positive_multiN}

The numerical results in the main text were obtained for $\eta=0$ and a
single population size. To determine whether moment equivalence depends on
these choices, we consider normalized exposure with $\eta=1$ and repeat the
calculations for $N=50,100,200,$ and $500$. To distinguish the distributions
used here from $Q_A$ and $Q_B$ in the main text, we denote the present pair
by $\widetilde{Q}_A$ and $\widetilde{Q}_B$. Their infection rates were
computed separately from the microscopic sums over group size and
composition, rather than from the reduced generator. The resulting overlap
therefore provides a direct numerical test of the moment-sufficiency theorem
in Subsec.~\ref{sec:exact_reduction}.

\begin{figure*}[tb]
    \centering
    \includegraphics[width=0.8\linewidth]
    {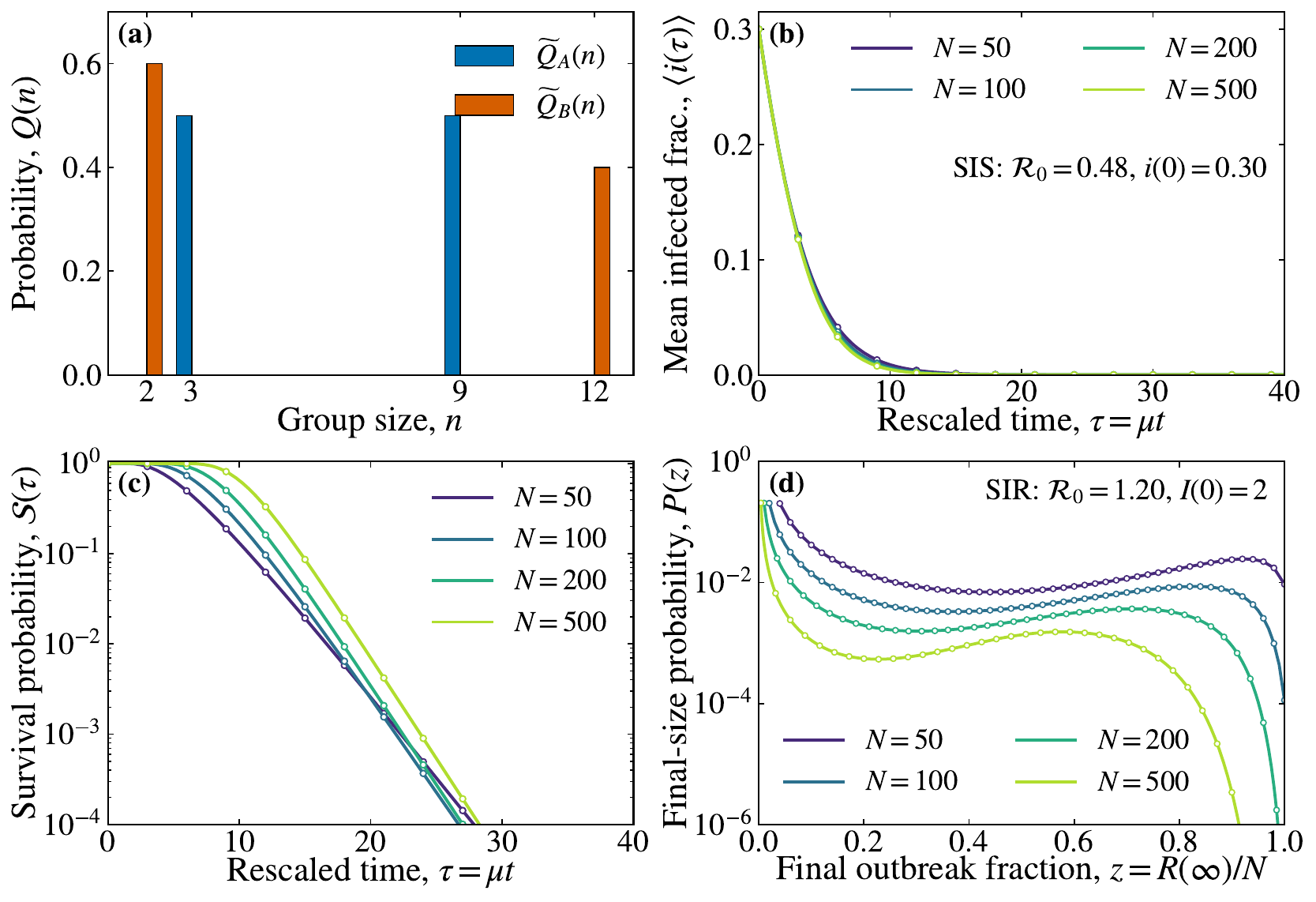}
    \caption{Moment equivalence under normalized exposure, $\eta=1$, for a
    quadratic kernel with $q=2$ and several population sizes.
    (a) Participant-view distributions
    $\widetilde{Q}_A(n)=\tfrac{1}{2}\delta_{n,3}
    +\tfrac{1}{2}\delta_{n,9}$ and
    $\widetilde{Q}_B(n)=\tfrac{3}{5}\delta_{n,2}
    +\tfrac{2}{5}\delta_{n,12}$. The two distributions have disjoint
    supports and satisfy
    $M_1^{(1)}[\widetilde{Q}_A]
    =M_1^{(1)}[\widetilde{Q}_B]=1$ and
    $M_2^{(1)}[\widetilde{Q}_A]
    =M_2^{(1)}[\widetilde{Q}_B]=4$, but differ at the next order, with
    $M_3^{(1)}[\widetilde{Q}_A]=21$ and
    $M_3^{(1)}[\widetilde{Q}_B]=36$.
    (b) Mean SIS prevalence and (c) survival probability
    $\mathcal{S}(\tau)=\Pr(\mu T_{\mathrm{ext}}>\tau)$ for
    $\mathcal{R}_0=0.48$, $i(0)=0.30$, and
    $N=50,100,200,$ and $500$.
    (d) Exact SIR final outbreak-size distributions,
    $z=R(\infty)/N$, for $\mathcal{R}_0=1.20$, $I(0)=2$, and the same
    population sizes. In panels (b)--(d), colors distinguish the population
    sizes, while solid lines and open circles represent
    $\widetilde{Q}_A$ and $\widetilde{Q}_B$, respectively. The open circles
    serve only as markers on the exact solutions. In panel (d), $P(z)$ is
    a discrete probability mass, and the lines are included as visual
    guides. The overlap obtained for every $N$ shows that dynamical
    equivalence is not restricted to $\eta=0$ or to a particular population
    size.}
    \label{fig:eta1_multiN_equivalence}
\end{figure*}

For $\eta=1$, the first generalized factorial moment satisfies
$M_1^{(1)}[Q]=1$ for every normalized distribution $Q(n)$, whereas
$M_2^{(1)}[Q]=\sum_n Q(n)(n-2)$. The two distributions shown in
Fig.~\ref{fig:eta1_multiN_equivalence}(a) have the same mean group size,
$\langle n\rangle_{\widetilde{Q}_A}
=\langle n\rangle_{\widetilde{Q}_B}=6$, and therefore both yield
$M_2^{(1)}=4$. Their agreement ends at third order, with
$M_3^{(1)}[\widetilde{Q}_A]=21$ and
$M_3^{(1)}[\widetilde{Q}_B]=36$. Because the quadratic kernel accesses only
the first two moments, the aggregate transition rates and Markov generators
of the two distributions are identical for every $N\geq12$.
Figures~\ref{fig:eta1_multiN_equivalence}(b)--\ref{fig:eta1_multiN_equivalence}(d)
show that changing the population size modifies the prevalence trajectory,
the SIS extinction statistics, and the SIR final outbreak-size distribution,
but does not separate the two moment-matched processes at a fixed $N$.
Across all population sizes considered, the maximum difference between the
infection rates generated by $\widetilde{Q}_A$ and
$\widetilde{Q}_B$ does not exceed $1.4\times10^{-15}$, while the maximum
differences in the SIS state distribution and the SIR final outbreak-size
distribution remain below $1.9\times10^{-14}$ and
$8.4\times10^{-17}$, respectively. These discrepancies are at the level of
floating-point roundoff and are consistent with exact generator equality.
The calculation provides a representative finite-population test for
$\eta=1$; for other values of $\eta$, the same equivalence holds whenever
the corresponding generalized factorial moments satisfy the matching
condition in Theorem~1.

\section*{REFERENCES}
\nocite{*}
\bibliography{aipsamp}% Produces the bibliography via BibTeX.

\end{document}